\documentclass[fleqn,usenatbib]{mnras}

\usepackage{newtxtext,newtxmath}

\usepackage[T1]{fontenc}
\usepackage{ae,aecompl}

\usepackage{graphicx}	
\usepackage{amsmath}	
\usepackage{multirow}
\usepackage{multicol}

\title[Attention for radio galaxy classification]{Attention-gating for improved radio galaxy classification}

\author[M. Bowles et al.]{
Micah Bowles,$^{1}$\thanks{E-mail: micah.bowles@postgrad.manchester.ac.uk (MB)}
Anna M. M. Scaife$^{1,2}$,
Fiona Porter$^{1}$,
Hongming Tang$^{1}$,
and 
David J. Bastien$^{1,3}$.
\\
$^{1}$Jodrell Bank Centre for Astrophysics, Department of Physics \& Astronomy, University of Manchester, Oxford Road, Manchester M13 9PL, UK\\
$^{2}$The Alan Turing Institute, Euston Road, London NW1 2DB, UK\\
$^{3}$Square Kilometre Array Organisation, Jodrell Bank Observatory, Macclesfield, SK11 9FT, UK
}

\date{Accepted 2020 December 18. Received 2020 December 18; in original form 2020 October 16}

\pubyear{2021}

\begin{document}
\label{firstpage}
\pagerange{\pageref{firstpage}--\pageref{lastpage}}
\maketitle

\begin{abstract}
In this work we introduce attention as a state of the art mechanism for classification of radio galaxies using convolutional neural networks. We present an attention-based model that performs on par with previous classifiers while using more than 50\% fewer parameters than the next smallest classic CNN application in this field. We demonstrate quantitatively how the selection of normalisation and aggregation methods used in attention-gating can affect the output of individual models, and show that the resulting attention maps can be used to interpret the classification choices made by the model. We observe that the salient regions identified by the our model align well with the regions an expert human classifier would attend to make equivalent classifications. We show that while the selection of normalisation and aggregation may only minimally affect the performance of individual models, it can significantly affect the interpretability of the respective attention maps and by selecting a model which aligns well with how astronomers classify radio sources by eye, a user can employ the model in a more effective manner.
\end{abstract}

\begin{keywords}
radio continuum: galaxies -- methods: statistical -- techniques: image processing
\end{keywords}



\section{Introduction}
\label{sec:intro}

As astronomers collect larger and larger volumes of data, machine learning techniques are becoming increasingly prevalent in astronomical analysis. Examples include the use of Support Vector Machines \citep[SVMs; see e.g.][]{svmreview} to study coronal mass ejections \citep{Qu2006}, gravitational lenses \citep{hartleysvm}, and for classifying astronomical objects detected in \textit{GAIA} data release 2 \citep{Bai2018, Prusti2016}, and random forests \citep[see e.g.][]{loupethesis} for the classification of data from SDSS data release 15 \citep{clarkesdss, sdssdr15}. Neural networks in astronomy were reviewed as early as \citet{Miller1993}, and have been present ever since, for example in \citet{Lahav1996} for galaxy classification and as recently as \cite{Das2019} to estimate spectroscopic mass, age and distance for red giant stars.

In radio astronomy a massive increase in data volume is driving the adoption of machine learning methodologies and automation. This is due to the range of new instruments that have recently come online, including the Low-Frequency Array \citep[LOFAR;][]{VanHaarlem2013}, the Murchison Widefield Array \citep[MWA;][]{Beardsley2019}, the MeerKAT telescope \citep{Jarvis2016}, and the Australian SKA Pathfinder (ASKAP) telescope \citep{Johnston2008}. For these instruments a natural solution has been to automate the data processing stages as much as possible, including classification of sources.

For example, the first fully public LOFAR Two-Metre Sky Survey \citep[LOTSS;][]{Shimwell2019} data release was mapped using a fully automated calibration and imaging process. This first data release covers only 2\,\% of the eventual coverage (424 square degrees) and catalogues 325,649 sources.  An object classification pipeline for LOTSS was developed and employed by \cite{Mingo2019RevisitingLoTSS}, who identified 5805 classifiable sources with active galactic nuclei (AGN) in LOTSS data release 1. This  pipeline split  the  identified AGN  into  Fanaroff-Riley Class~I (FRI) and Fanaroff-Riley Class~II \citep[FRII;][]{FR1974}. This  morphological classification of radio loud AGNs separates `edge darkened' FRI galaxies from `edge brightened' FRII galaxies.
 
 Over the years the morphological dichotomy of the FR classification has been widely investigated through both observations and simulations of complex sources \citep[e.g.][]{Smithand2019,Schoenmakers2000,Mahatma2019}. Various suggestions of additions and alternate classifications have also been made. A suggestion for the addition of an `FR0' class was made in \cite{Baldi2015} for compact radio sources and a number of studies have investigated these sources, their properties and their relation to the prevailing FRI and FRII source classes  \citep{Baldi2016,Baldi2019,Torresi2018,Capetti2020}. From the LOTSS sample, \cite{Mingo2019RevisitingLoTSS} found that the traditional dichotomy of the FR classification was not sufficient to describe the complex morphologies of the detected sources, and they additionally classified `low luminosity FRII' sources, which extended to three orders of magnitude below the luminosity break observed to accompany the morphological distinction of FRI/II sources. Furthermore, various sub-classes of the FR scheme are commonly employed to label specific structures commonly found in radio galaxies. Both \cite{MiraghaeiBest17} and \cite{Mingo2019RevisitingLoTSS} include small samples of sources which do not clearly adhere to the FR dichotomy and it is certain that deeper and wider radio surveys will open up even more detailed views of such source morphologies. Additionally, a hybrid morphology, where a given source presents FRI properties on one side of its core and FRII properties on the other, is also widely recognised \citep{Gopal-Krishna2000, Kapinska2017, Seymour2020}.

In spite of these known issues, the FRI/II classification scheme persists, with recent research still seeking to understand through which processes and under what conditions these sources evolve \cite[e.g.][]{MiraghaeiBest17,InesonCroston15}. Large scale samples of FRI/II classifications are used to evaluate theoretical and simulated population models \citep[e.g.][]{Godfrey2017,Hardcastle2018}, model the evolution of radio-AGN \citep[e.g.][]{Best2014}, and support the development of a unified AGN model \citep{Netzer2015}.
Further uses for these radio sources are summarised by \cite{Hardcastle2020}. These include uses within cosmic magnetism \citep[e.g.][]{Bonafede2010,Govoni2010,OSullivan2019} and cosmology \citep[e.g.][]{Raccanelli2012}. Regardless of the field using these sources, if the sources are not confidently classified, the application suffers. 

It is therefore perhaps unsurprising that an increasing number of works in radio astronomy have been developing machine learning approaches to classify radio galaxies in the Fanaroff-Riley scheme \citep[e.g.][]{Aniyan2017,Lukic2019,Tang2019a,Ma2019}. Since this is primarily a morphological classification these approaches have focused on the use of convolutional neural networks (CNNs); however, the computational cost of training typical commercial CNNs with 10s to 100s of millions of parameters is not insignificant and furthermore, for problems with comparatively small training data sets such networks can easily result in over-fitting.

For example, the classic CNN architecture VGG16 \citep{simonyan2014deep}, is a deep convolutional neural network developed by the Visual Geometry Group at Oxford. 
VGG16 is able to fit complex visual tasks and can discriminate the $1000$ classes of the ImageNet data set \citep{imagenet_cvpr09} with a top-1 value-error (i.\,e. the percentage of top predictions that are incorrect) of approximately 25\,\%. At the time of its publication this was the highest performing model for the ImageNet data set and VGG16 became popular due to its structural simplicity and performance. However, with 138\,M learnable parameters, VGG16 requires extremely large training data sets such as Imagenet, containing $\sim15$\,million images, to realise its purpose.

In radio astronomy, as with other domain specific applications of deep learning where labelled training data sets are often significantly smaller, it is beneficial to adjust the structure and size of CNNs. One such example is the network proposed and evaluated in \cite{Tang2019a} which was trained on FRDEEP, a data set of 600 classified radio sources. With many fewer degrees of freedom, this problem does not require a network as deep as VGG16 and the implemented model has only 250\,k parameters in comparison to the 138\,M of VGG16.

Similarly, the AG-Sononet architecture \citep{SononetAGSchlemperOktay} was first introduced for classification in medical imaging, specifically to classify sonogram images. Although AG-SonoNet has the same number of convolution operations in its main body as VGG16, it does not use VGG16's fully connected layers, which constitute 119\,M (90\,\%) of that network's 138\,M parameters, and AG-Sononet therefore contains only 696\,k parameters. In order to achieve this reduction in parameter volume AG-Sononet employs an attention gating mechanism to perform classification rather than a fully-connected network.

Machine learning applications of attention are increasingly used to improve both the performance and interpretation of machine learning models \citep{Ba2014,StollengaMGS14,Bahdanau2015NeuralMT,Xu2015,Chen2017,Jetley2018}.
These applications are analogous to the biological concept of attention \citep{LindsayAttentionReview2020,Itti2001,Zhou2011}, whereby the visual system prioritises the most salient features in an image, i.e. the feature containing the most information pertinent to the context.

In this work we introduce the concept of attention-gating to radio galaxy classification. We demonstrate that attention-gated networks can provide equivalent model performance to existing CNN-based radio galaxy classification whilst using significantly fewer trainable parameters. Furthermore, we demonstrate that the attention maps produced by these networks can be used to aid the interpretability of such machine learning models for astronomical applications. The structure of this paper is as follows: in Section~\ref{sec:attention} we introduce the attention mechanism for convolutional neural networks; in Section~\ref{sec:arch} we describe the network architecture deployed in this work and the implementation of the attention gates themselves; in Section~\ref{sec:data} we give an overview of the radio astronomy data sets used for this work; in Section~\ref{sec:performance} we provide details of the model performance with reference to alternatives in the literature and justify our choice of normalisation and aggregation method for the attention gates; in Section~\ref{sec:distr_attention} we examine the average attention distribution as a function of target class across the data set and discuss its interpretation; in Section~\ref{sec:individual_sources} we consider how attention distributions may inform a user about mis-classifications in a data set; and in Section~\ref{sec:conclusion} we summarise and draw our conclusions.

\section{Attention}
\label{sec:attention} 

There are two clear approaches to attention in machine learning: hard spatial attention and soft spatial feature attention. These two approaches have clear alignments to overt and covert visual attention in the biological sense, respectively.

\subsection{Hard vs. Soft Spatial Attention}

When a ML algorithm outputs multiple sequential outputs based on individual sequential inputs selected by the model, this is considered hard attention. This has become common in natural language processing \citep{Galassi2019CriticalReviewNLPAttention}. For example, 
consider the input $\mathbf{x}=(x_1, ... , x_T)^\intercal$, where each element of $\mathbf{x}$ refers to an English word, which is to be translated to an output $\mathbf{y}$ in another language. To do this, an encoder-decoder network is used where the encoder is a Bidirectional Recurrent Neural Network (BiRNN) and the decoder is comprised of an attention function and a Recurrent Neural Network (RNN). The BiRNN encodes the input to \textit{annotations}, $\mathbf{H}$, as:
\begin{equation}
    \mathbf{H} = (\mathbf{h}_1, ... , \mathbf{h}_T) = \text{BiRNN}(x_1, ... , x_T).
\end{equation}

The selected attention function uses these annotations to calculate a \textit{context vector}, $\mathbf{c}_t$, for the next position $t$ in the translated sentence as:
\begin{equation}
    \mathbf{c}_t = \text{Attention}(\mathbf{H},\mathbf{s}_{t-1}),
\end{equation}
where $\mathbf{s}_{i}$ are hidden states used by the final RNN to extract the most probable next word in the translated sentence given the input and the translated sentence output so far:
\begin{equation}
    p(y_t|y_1,...,y_{t-1},\mathbf{x}) = \text{RNN}(\mathbf{c}_t,\mathbf{s}_{t-1}).
\end{equation}

The attention operation itself is best explained in three steps: first, the annotations and hidden state of the previous output are passed into a function, $f$, which is typically a feedforward neural network \citep[e.g.][]{Bahdanau2015NeuralMT}, which creates a set of scalar values, $e_{ti}$, that score the relatedness of the inputs around $i$ to the outputs around $t$:
\begin{equation}
    e_{ti} = f(\mathbf{s}_{t-1},\mathbf{h}_i).
\end{equation}

This value is normalised across each of the inputs, using a softmax function, to create a set of weights $\alpha_{ti}$:
\begin{align}
    \alpha_{ti} &= \frac{\exp{e_{ti}}}{\sum^T_{j=1}\exp{e_{tj}}},
\end{align}
which are then used to scale the annotations and return the context vector:
\begin{equation}
    \mathbf{c}_t = \sum_i \alpha_{ti}\mathbf{h}_i = \text{Attention}(\mathbf{H},\mathbf{s}_{t-1}).
\end{equation}

This kind of hard attention has also been used for computer vision applications such as generating image captions for a given input by `translating' image regions \citep{Xu2015} and for classifying multiple objects in single images \citep{Ba2014}. Such hard attention selection for visual applications can be seen as being analogous to saccadian motion selecting salient regions with the fovea, and thus overt visual attention \citep[e.g.][]{Itti2001}.

Soft spatial attention, also referred to as feature attention, scales the representation according to spatial location equally across all feature maps and amplifies individual channels to scale certain features regardless of spatial location. \cite{Chen2017} used both soft spatial and feature attention for image captioning tasks, whereas \cite{StollengaMGS14} generated soft feature attention by assigning weights to each feature map according to the feature maps generated in a forward pass of an image through the network. 

The attention gates implemented in this work can be described as being \emph{soft trainable attention}, which defines both spatial and feature saliency maps used to attend an input. These were initially introduced in \cite{Jetley2018}, and were clarified and re-implemented in \cite{Schlemper2019}. 

\subsection{Attention Gates}
\label{sec:AttentionGates}

Conceptually, attention gates amount to filters that prioritise salient features and their respective spatial locations within a given input. Soft feature attention is enforced using [1$\times$1]-convolutions, which scale each feature (channel) of an input according to a learned weight. The attention weight is implemented as a normalised single-channel attention map which is a weighted sum of each of the salient features present in the image.

\begin{figure*}
    \centering
    \includegraphics[width=\linewidth]{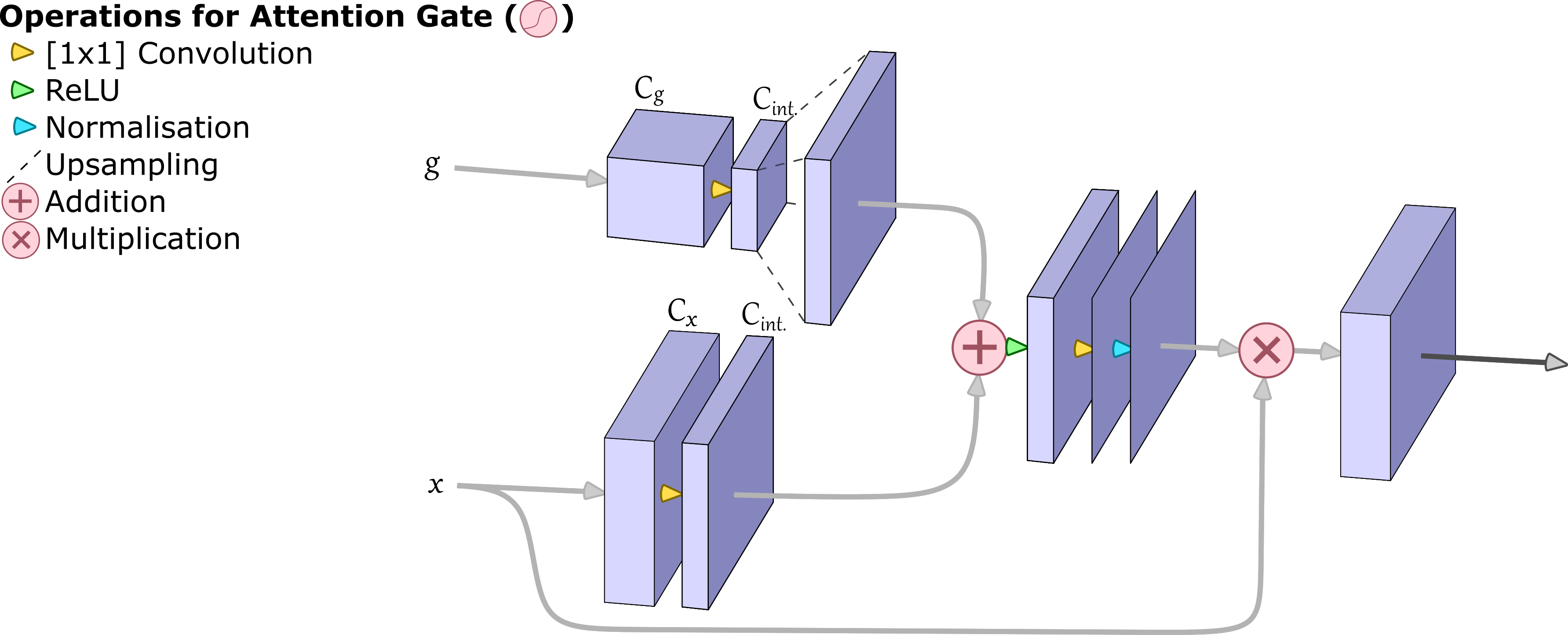}
    \caption{The attention gate implemented in this work, which learns soft feature and spatial visual attention (see Equations~\ref{eqn:AGOutput} and \ref{eqn:AGCompatability}).}
    \label{fig:AttentionGate}
\end{figure*}

More specifically, an input, $\mathbf{x}\in\mathbb{R}^{C_x\times H_x\times W_x}$, is attended using a compatibility score, $C(\mathbf{x},\mathbf{g}): \mathbb{R}^{C_x\times H_x\times W_x}\times\mathbb{R}^{C_g\times H_g\times W_g}\rightarrow\mathbb{R}^{H_x\times W_x}$, to generate an (attended) output, $\boldsymbol{\alpha}\in\mathbb{R}^{C_x\times H_x\times W_x}$, following:
\begin{equation}
    \alpha_{ijk} = \sigma_2[C(\mathbf{x},\mathbf{g})_{jk}] \cdot x_{ijk}, \label{eqn:AGOutput}
\end{equation}
with a normalisation $\sigma_2$ and the compatibility score, $C(\mathbf{x},\mathbf{g})$, calculated as:
\begin{equation}
    C(\mathbf{x},\mathbf{g}) = \left[\sigma_1(\mathbf{x}*\theta+(\mathbf{g}*\Phi)')\right]*\Psi, \label{eqn:AGCompatability}
\end{equation}
where the $*$ operator denotes the convolution operation; $\theta$, $\psi $ and $\Psi$ are [1$\times$1] convolutions; $\sigma_1$ is the ReLU non-linearity; and the prime in $(\mathbf{g}*\psi)'$ refers to the up-sampling required to match width and height dimensions to $(\mathbf{x}*\theta)$. This process is visualised in Figure~\ref{fig:AttentionGate}.

The [1$\times$1]-convolutions are chosen such that $(x*\theta)$ and $(g*\psi)$ share the same intermediate channel number as their channel dimension output, which allows for simple addition of the two tensors. Furthermore, $\Psi$ is a [1$\times$1]-convolution which takes this intermediate channel dimension and reduces it to the one channel width of the compatibility score. Once normalised this compatibility score becomes the \textit{attention map} (saliency map). 

\subsection{Aggregation Methods}
\label{sec:Aggregation}

Once attention gates have been applied within a CNN, the different feature maps are used to generate an output. While in conventional CNN architectures this is achieved using multiple fully connected layers, in the case of attention-gated networks the number of fully connected layers is minimised in order to increase the classifier's dependency on the attention gates themselves, the outputs from which are aggregated for classification.

Although in principle an aggregation method could be implemented in any way the user desires, in this work the four methods implemented by \cite{Schlemper2019} are considered.
In these methods, the output of the attention gates remains as $\alpha^n\in\mathbb{R}^{C_n\times H_n\times W_n}$ with a superscript $n$ to indicate which attention gate the output corresponds to. The output of each aggregation method is a vector with length equivalent to the number of classes $y\in\mathbb{R}^{n_{\text{classes}}}$, where in this work $n_{\text{classes}}=2$. The output prediction for each class is evaluated as $y_{\text{out}}=\text{Softmax}(y)$.

The proposed mechanism classifies on the summed value of each channel of the attention gate's output, meaning that the values $f_i^n$ as defined by Equation~\ref{eqn:feature_sum} are used to make the classification, such that
\begin{equation}
    f^n_i = \sum_{j,k}\alpha^n_{ijk}, \label{eqn:feature_sum}
\end{equation}
with $i\in[0,...,C_n]$ and $n\in[1,...,N]$ where $N$ is the selected number of attention gates. Logits are then constructed using one of the following four methods.

\subsubsection{Mean}
In this method the classification is made by taking the mean of multiple fully connected layers, each applied to the attention maps $f^n_i$:
\begin{equation}
    \mathbf{y} = \frac{1}{N}\sum_n \mathbf{W}_n \mathbf{f}^n + \mathbf{b}_n,\label{eqn:Agg_mode_Mean}
\end{equation}
where $\mathbf{W}_n\in\mathbb{R}^{2\times C_n}$ and $\mathbf{b}_n\in\mathbb{R}^2$ are learnable parameters.

\subsubsection{Concatenation}
In this method a classification is made on the concatenation of all of the feature maps:
\begin{equation}
    \mathbf{y} = \mathbf{W}\left(\begin{matrix}\mathbf{f}^1\\\mathbf{f}^2\\\mathbf{f}^3\end{matrix}\right)+\mathbf{b},\label{eqn:Agg_mode_Concatenation}
\end{equation}
where $\mathbf{W}\in\mathbb{R}^{2\times (C_1+C_2+C_3)}$ and $\mathbf{b}\in\mathbb{R}^2$ are learnable parameters.

\subsubsection{Deep Supervised}
The deep supervised method is an expanded version of the mean aggregation, where the final classification is an average of both individual classifications given in Equation~\ref{eqn:Agg_mode_Mean} and the concatenated classification given in Equation~\ref{eqn:Agg_mode_Concatenation}:
\begin{equation}
    \mathbf{y} = \frac{1}{N+1}\left[\mathbf{W}
    \left(
    \begin{matrix}\mathbf{f}^1 \\ \mathbf{f}^2 \\ \mathbf{f}^3\end{matrix}
    \right)
    +\mathbf{b} + \sum_n \mathbf{W}_n \mathbf{f}^n + \mathbf{b}_n\right],
\end{equation}
where $\mathbf{W}$ and $\mathbf{b}$ are defined as in the definition of the concatenation method, and $\mathbf{W}_n$ and $\mathbf{b}_n$ are defined as in the definition of the mean method.

\subsubsection{Fine Tuned}
The fine tuned method employs a single fully connected layer to classify on the classifications made on each individual attention gate:
\begin{equation}
    \mathbf{y} = \mathbf{W}_{ft}
    \left(\begin{matrix}
            \mathbf{W}_1\mathbf{f}^1+\mathbf{b}_1 \\
            \mathbf{W}_2\mathbf{f}^2+\mathbf{b}_2 \\
            \mathbf{W}_3\mathbf{f}^3+\mathbf{b}_3
        \end{matrix}\right) + \mathbf{b}_{ft},
\end{equation}
where $\mathbf{W}_{ft}\in\mathbb{R}^{2\times6}$, $\mathbf{b}_{ft}\in\mathbb{R}^2$,
$\mathbf{W}_n\in\mathbb{R}^{2\times C_n}$ and $\mathbf{b}_n\in\mathbb{R}^{2}$ are all learnable parameters.

\section{Network Architecture}
\label{sec:arch} 

The network in this work is inspired by the SonoNet architecture \citep{Schlemper2019}, which in turn was inspired by the VGG-16 architecture \citep{simonyan2014deep}. We alter the base SonoNet architecture by removing the final pooling layer and all subsequent convolutional layers. This was done to reduce the complexity of the network and prevent over-fitting. Over-fitting is a serious problem when using deep networks and can be combated using validation methods. In this work, early stopping is implemented as the validation method of choice. As such a portion of the training data is reserved for validation. This model's loss on this validation set is recorded throughout training, with the final output being the model which achieved the minimal validation loss throughout training.

Validation tests showed that the original structure of the Sononet architecture applied to radio astronomy images quickly resulted in over-fitting of the data and the network was truncated in response. Similarly, if there are not enough learnable parameters in a network, the trained model may not able to differentiate between target classes and will either predict randomly or predict all sources to belong to a single class. For example, when the structure from \cite{Tang2019a} was adapted to attention gates, this over-fitting occurs immediately. The largest difference here, is that the fully connected layers, which contain 94\,\% of the original parameters, are removed for the attention gated implementation. Although the original performs well, the adapted model contains too few parameters and is not able to fit to the data correctly.

\begin{figure*}
    \centering
    \includegraphics[width=\linewidth]{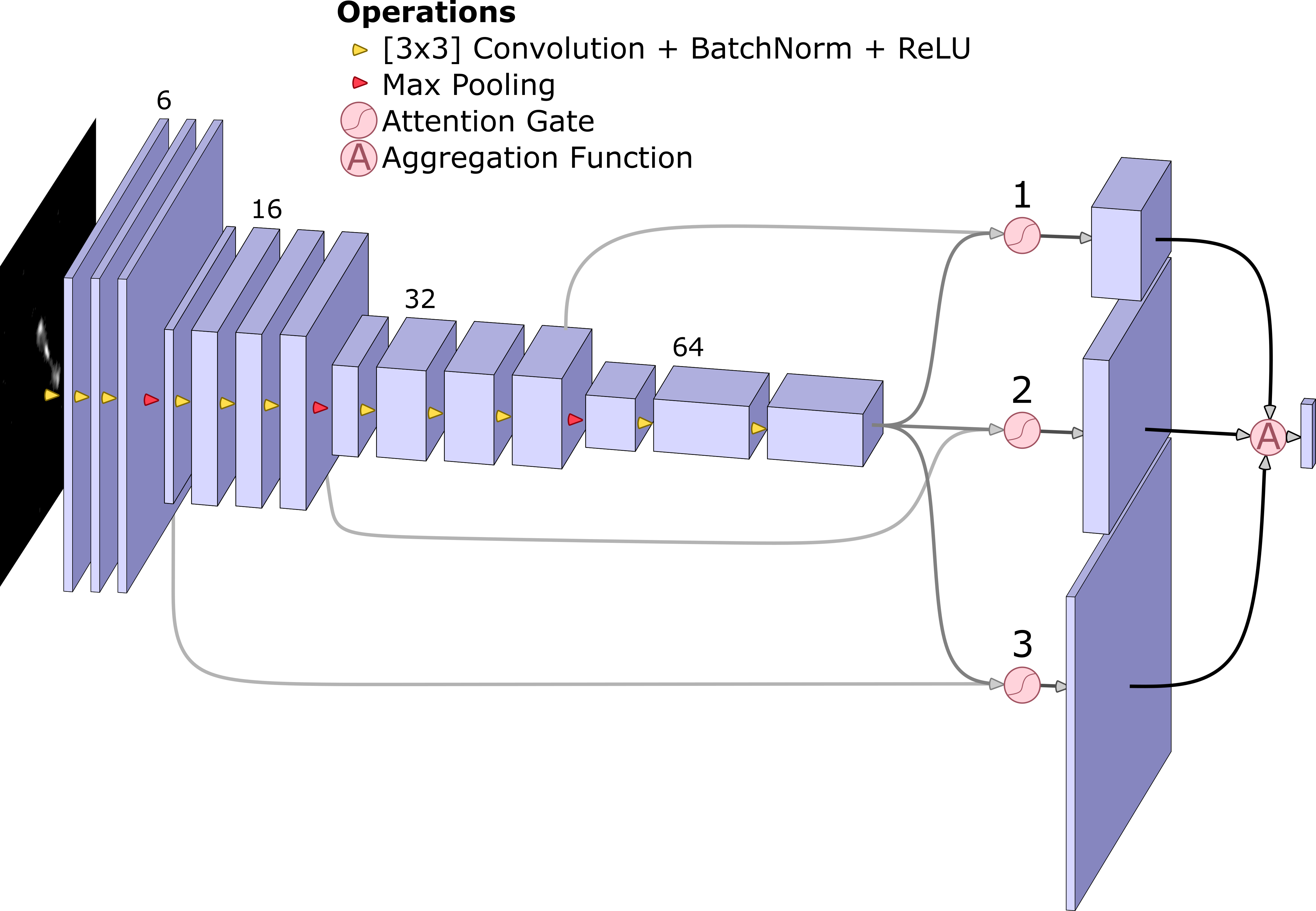}
    \caption{The primary network architecture implemented in this work. The attention gates are added in numbering order shown here, according to how many attention gates are requested. Channel width is given as a digit above each altered layer.}
    \label{fig:RadGalNet_Architecture}
\end{figure*}

A detailed summary of the primary network implemented in this work and its parameters is given in Table~\ref{tab:RadGalNet} and depicted in Figure~\ref{fig:RadGalNet_Architecture}. In the following sections we refer to the architecture implemented here as the \emph{AG-CNN}.
\begin{table*}
    \centering
    \begin{tabular}{|l |r r r|r|}
        \hline
        \multirow{2}{*}{Operation Block} & \multicolumn{3}{c|}{Output Shape} & Number of \\ 
                                         & Channel & Width & Height        &  Parameters\\ \hline\hline
                                          &1&150&150  &0      \\
        $[3\times3]$ Conv. + ReLU + BNorm &6&150&150  &72     \\
        $[3\times3]$ Conv. + ReLU + BNorm &6&150&150  &342    \\
        $[3\times3]$ Conv. + ReLU + BNorm &6&150&150  &342    \\
        Max Pooling                       &6&75&75    &0      \\
        $[3\times3]$ Conv. + ReLU + BNorm &16&75&75   &912    \\
        $[3\times3]$ Conv. + ReLU + BNorm &16&75&75   &2,352  \\
        $[3\times3]$ Conv. + ReLU + BNorm &16&75&75   &2,352  \\
        Max Pooling                       &16&37&37   &0      \\
        $[3\times3]$ Conv. + ReLU + BNorm &32&37&37   &4,704  \\
        $[3\times3]$ Conv. + ReLU + BNorm &32&37&37   &9,312  \\
        $[3\times3]$ Conv. + ReLU + BNorm &32&37&37   &9,312  \\
        Max Pooling                       &32&18&18   &0      \\
        $[3\times3]$ Conv. + ReLU + BNorm &64&18&18   &18,624 \\
        $[3\times3]$ Conv. + ReLU + BNorm &64&18&18   &37,056 \\
        Attention Gate 1                  &32&37&37   &6,209  \\
        Attention Gate 2                  &16&75&75   &5,185  \\
        Attention Gate 3                  &6&150&150  &4,545  \\
        Sum Across Height and Width       & & &54     &0      \\
        Dropout                           & & &54     &0      \\
        Aggregation Function              & & &2      &14     \\
        \hline\hline
        Total Parameters: &\multicolumn{4}{r|}{101,447} \\
        \hline
    \end{tabular}
    \caption{\label{tab:RadGalNet} A summary of the primary network implemented in this work and visualised in Figure \ref{fig:RadGalNet_Architecture}, its parameters and the size of each of the feature maps. This is a valid for implementations with the fine tuned aggregation, any normalisation and three attention gates.}
\end{table*}

\subsection{Attention Implementation}
\label{sec:attention_implementation}

As described in Section~\ref{sec:AttentionGates}, the form of the attention gates used in this work is equivalent to that in \cite{Schlemper2019}. However, the implementation of these gates within the network itself is not the same. In the architecture adopted here the attention gates are implemented such that they are the only input on which the network makes a classification, thus guaranteeing that the attention gate outputs are used to make the final classification.

SonoNet uses two attention gates and additionally uses the final feature map of the convolutional layers as an input for the aggregation method. In principle, aggregation of these results allows a classification to be made based purely on the final feature map, and does not (theoretically) require information from the attention gates to be used at all. In this work we implement up to three attention gates and only classify on the output of the attention gate(s) themselves. This is more similar to the initial learned CNN attention gate implementation \citep{Jetley2018}, where it was noted that classification under such a restriction is more consistent with the original concept of attention as practiced in NLP.

Table~\ref{tab:AG_parameter_summary} summarises the parameters of Attention Gate 1 as an example of how the parameters align within each attention gate; see also Figure~\ref{fig:AttentionGate}. In this work, as well as exploring each of the aggregation methods defined in Section~\ref{sec:Aggregation}, we also consider four different attention normalisation functions for the attention gates, denoted $\sigma_2$ in Equation~\ref{eqn:AGOutput}. These are summarised in Table~\ref{tab:norms}. Dropout is applied to the summed channel (feature) values output from the attention gates, prior to aggregation, see Equation~\ref{eqn:feature_sum}.
\begin{table}
    \centering
    \begin{tabular}{|l |r r r|r|}
        \hline
        \multirow{2}{*}{Operation Block} & \multicolumn{3}{c|}{Output Shape} & Number of \\ 
                                         & Channel & Width & Height        &  Parameters\\ \hline\hline
        Input               &32&37&37   &9,312  \\
        Global Input        &64&18&18   &37,056 \\ \hline\hline
        $[3\times3]$ Conv.  &64&37&37   &2,048  \\
        $[3\times3]$ Conv.  &64&18&18   &4,096  \\
        Upsample            &64&37&37   &0      \\
        $[3\times3]$ Conv.  &1&37&37    &65     \\
        Normalise Att. Map  &1&37&37    &0      \\
        Input $\times$ Att. Map &32&37&37 &0    \\
        \hline\hline
        Attention Gate 1: &\multicolumn{4}{r|}{6,209} \\
        \hline
    \end{tabular}
    \caption{Summary of attention gate 1, its parameters and the size of each of the feature maps. These values are valid for any of the implemented normalisations as they have no learnable parameters.}
    \label{tab:AG_parameter_summary}
\end{table}
\begin{table}
    \centering
    \begin{tabular}{cl|c}
        \hline
        & Function & Functional form \\\hline
        (i) & Softmax &  $\sigma_2(x) = {\rm e}^{x}/\sum{{\rm e}^{x}}$\\
        (ii) & Sigmoid &  $\sigma_2(x) = \left(1+{\rm e}^{-x}\right)^{-1}$\\
        (iii) & Range & $\sigma_2(x) = (x-\text{min}(x))/(\text{max}(x)-\text{min}(x))$\\
        (iv) & Standardisation & $\sigma_2(x) = (x-\mu)/\sigma$ \\\hline
    \end{tabular}
    \caption{Attention normalisation functions. Here min and max are defined as functions which return the minimal and maximal values of their respective inputs, and $\mu$ and $\sigma$ are the chosen mean and standard deviation of the input, which are selected to be $\mu=0.5$ and $\sigma=0.25$.}
    \label{tab:norms}
\end{table}

The network implementation is made in such a way that not all attention gate outputs are necessary for classification. Where classification is made using outputs from fewer than the maximum number of three attention gates the models include gates $1 - 3$, as shown in Figure~\ref{fig:RadGalNet_Architecture}, sequentially, i.e. \emph{2 attention gates} implies that gates 1 and 2 are used. In the case where no attention gates are included an additional max-pooling layer followed by a single fully-connected layer is used for classification.

Models are trained over 100 epochs, using the Adam optimiser and an initial learning rate of $5\cdot10^{-5}$ (adapted only once to enable the given model to train). The models trained on the MiraBest \citep{Porter2020} data set for this work took an average of $8\text{\,hrs\,} 18\text{\,min\,} \pm \text{\,}1\text{\,hr\,} 20\text{\,min}$ on an 8GB Nvidia RTX\,2080 GPU.
Depending on the hardware used and the optimisation process, i\,e. how many times the intermittent models are saved to disk, the training time can vary significantly. Further training time dependencies include the model's hyperparameters, e.\,g. the choice of optimiser, learning rate, training epochs, data set etc. Given a pre-processed image, the trained model can output its predicted label in $\sim0.6\,\text{ms}$
, as measured over 1000 augmentations of the MiraBest test set.

\section{Data}
\label{sec:data}

\subsection{Data Sets}
\label{sec:datasets}

The two data sets used in this work both
use image data from the VLA FIRST survey \citep{Becker1995}, with the number (and sources) of labels derived from different sources. The data sets themselves are composed as follows:

\subsubsection{FR-DEEP}
The FR-DEEP data set was first presented in its entirety in \citet{Tang2019a}. 
The labels for the FR-DEEP data set were taken from the CONFIG \citep{Gendre2008,Gendre2010} and FRICAT \citep{Capetti2017FRICAT:Galaxies} catalogues, where sources were visually classified by their expert authors. \citet{Tang2019a} selected a subset from those catalogues to include only sources that were denoted as confidently classified. In this work the FR-DEEP-F subset is used, which contains source images from the VLA FIRST survey at 1.4\,GHz \citep{Becker1995}. The data set contains 264 images labelled FRI and 336 sources labelled FRII.

\subsubsection{MiraBest}
The MiraBest data set \citep{Porter2020} is comprised of 1,256 radio galaxies with labels assigned using the catalogue of \citet{MiraghaeiBest17}, where labels were assigned using visual inspection.
\begin{table*}
    \begin{center}
    \begin{tabular}{|c|c|c|c|c|c|c|}
        \hline
        Label Used & No. & Class & Confidence & Morphology & No. & MiraBest Label \\ 
        \hline\hline
        \multirow{5}{*}{0} & \multirow{5}{*}{591} & \multirow{5}{*}{FRI} & \multirow{3}{*}{Certain} & Standard & 339 & 0 \\ \cline{5-7}
         &  &  &  & Wide-Angle Tailed & 49 & 1 \\ \cline{5-7}
         &  &  &  & Head-Tail & 9 & 2 \\ \cline{4-7}
         &  &  & \multirow{2}{*}{Uncertain} & Standard & 191 & 3 \\ \cline{5-7}
         &  &  &  & Wide-Angle Tailed & 3 & 4 \\  \hline\hline
        \multirow{3}{*}{1} & \multirow{3}{*}{631} & \multirow{3}{*}{FRII} & \multirow{2}{*}{Certain} & Standard & 432 & 5 \\ \cline{5-7}
         &  &  &  & Double-Double & 4 & 6 \\ \cline{4-7}
         &  &  & Uncertain & Standard & 195 & 7 \\  \hline\hline
        \multirow{2}{*}{NA} & \multirow{2}{*}{34} & \multirow{2}{*}{Hybrid} & Certain & NA & 19 & 8 \\ \cline{4-7}
         &  &  & Uncertain & NA & 15 & 9 \\ \hline
    \end{tabular}
    \end{center}
\caption{MiraBest data set summary. The original data set labels (MiraBest Label; \citet{Porter2020}) are shown in relation to the labels used in this work (Label). Hybrid sources are not included in this work, and therefore have no label assigned to them.} 
\label{tab:MiraBestClasses}
\end{table*}

Although the MiraBest data set also contains sub-classifications of FR sources, in this work we use it only for binary FRI/FRII classification. Table~\ref{tab:MiraBestClasses} shows the labels used in this work in relation to the more detailed labels provided by the MiraBest data set itself.
We do not include objects classified as \emph{Hybrid}. Furthermore the MiraBest data set flags individual objects as confidently classified (Certain) or unconfidently classified (Uncertain) depending on how much human interpretation was required to label a specific object, as described in \citet{MiraghaeiBest17}. For the remainder of this work, we refer to the full Certain+Uncertain data set as \emph{MiraBest} and to the Certain subset as \emph{ MiraBest$^{\ast}$}. Unless otherwise stated, the models in this work are trained and evaluated on MiraBest.

\subsection{Pre-Processing}
\label{sec:pre-processing}

FR-DEEP's pre-processing, as described in \cite{Tang2019a}, is the same pre-processing as was followed to create the MiraBest data set. The extracted images are processed in three stages before data augmentation is applied. 

First the image is clipped: the image pixel values are set to zero if their value is below a threshold of three times the root mean squared (RMS) signal of the local noise which was determined by a pixel histogram fit for each source image, this approach may clip out diffuse, low-surface brightness emission but was selected to align with \citet{Aniyan2017} who suggest this is the best clipping level for radio galaxy classification. This removes most artefacts and leaves behind cleaner images with clear sources. Any future inputs to the model should be treated in the same manner to allow the model to classify according to the features it has learned, and not the noise which the original image may contain.

The second step is to clip the image size to 150 by 150 pixels, i.\,e. $270^{\prime\prime}$ by $270^{\prime\prime}$ for FIRST where each pixel corresponds to $1.8^{\prime\prime}$ by $1.8^{\prime\prime}$. This is to standardise the size of the image and to provide the model with an image which ideally only contains the source of interest. However, by visual inspection we estimate that the clipped FR-DEEP-F and MiraBest data sets contain $\sim 1.2$ and $1.4$ sources per image, respectively.

Finally, the image is then normalised as:
\begin{equation}
    \text{Final} = 255\cdot\frac{\text{Img} - \text{min}(\text{Img})}{\text{max}(\text{Img})-\text{min}({\text{Img}})},
\label{eqn:pre_processing_norm}
\end{equation}
where `Final' is the normalised image, `Img' is the original image and `min' and `max' are functions which return the single minimal and maximal values of their input respectively. The steps in this pre-processing are illustrated in Figure~\ref{fig:data_processing_full_example}.

\subsection{Data Augmentation}
\label{sec:data_augmentation}

Neither the FRDEEP nor MiraBest data sets are sufficiently abundant that all angles and possible positions of radio galaxies are represented within the data. Human observers easily recognise that there is no class difference between an FRI galaxy and the same galaxy rotated by $180^{\circ}$, however, the volume of data does not allow for the ML optimisation to `learn' this invariance. To allow for this, a commonly implemented process is \textit{data augmentation}, which can be defined as the process by which the volume of training, validation and testing data is artificially inflated. 

\begin{figure*}
    \centering
    \includegraphics[width=\linewidth]{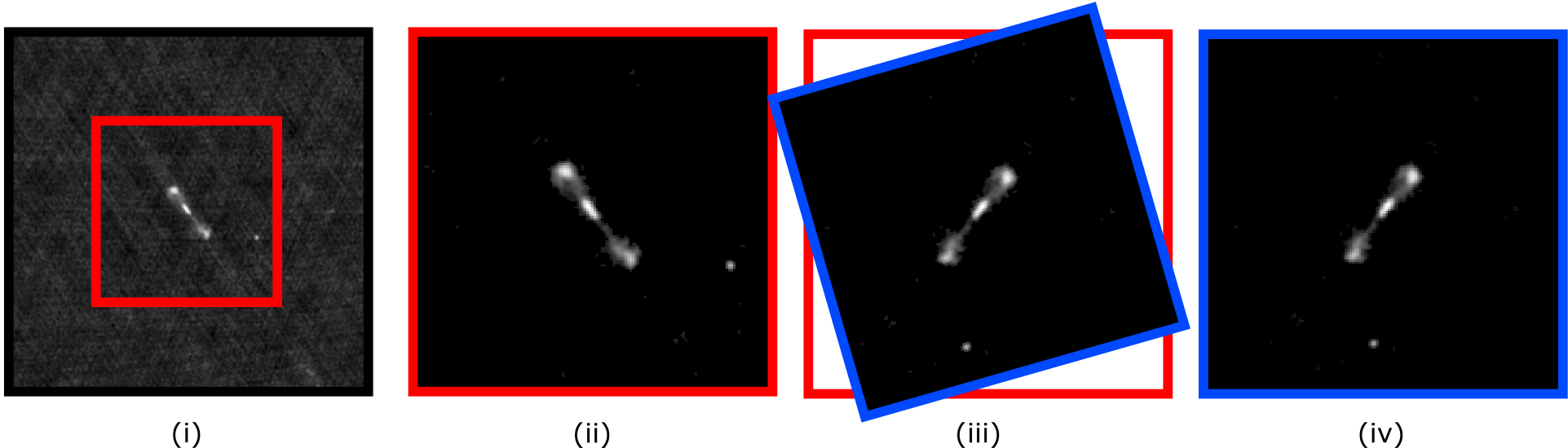}
    \caption{Illustration of the pre-processing of a source from the MiraBest test set. The individual panels present: (i) The original image extracted from the FIRST survey with a bounding box showing the applied crop. (ii) The cropped and sigma clipped image supplied by MiraBest. (iii) The equivalent data augmented image. (iv) A final augmented image with zero padding in remaining unfilled regions.}
    \label{fig:data_processing_full_example}
\end{figure*}

In the case of radio galaxies, for instance, the classification of a given source is independent of the scale, position or orientation of the sources within the image. Therefore, such invariances can be introduced during training by applying image transformations that enable the model to generalise to sources at scales, positions or orientations it would otherwise be unfamiliar with. Figure~\ref{fig:data_processing_full_example} shows the effects of the transformations used in this work. From the perspective of parameter optimisation, the model is now able to sample the parameter space in positions which are valid, but would otherwise remain unsampled due to the limited size of the data set. We note that the transformations that are applied during data augmentation should not be selected without careful consideration of the problem itself as not all data sets are rotationally invariant.

The transformation function of a given input, $F(x)$, implemented in this work can be summarised as:
\begin{equation}
    F(x) = R_{[-180,180]}\circ T_{[\pm 2\text{px},\pm2px]}\circ S_{[0.9,1.1]}(x),
\end{equation}
where $x\in[0,1]^{150\times 150}$ is the processed input image, $S$ is a scaling translation which (from the centre of the image) randomly scales the input by a factor in the range $[0.9, 1.1]$, $T$ is randomly applied translation operation which translates the image a number of pixels in $[-2,2]\in\mathbb{N}$ both vertically and horizontally, where the pixel distance to translate is independently selected for each axis, i.\,e. $T_{[2,-1]}(x)$ is a valid translation, and $R$ applies a rotation to the image around an angle randomly selected within the range $[-180,180]$ using bilinear interpolation.

Assuming that at least three scales should be applied by $S$, and knowing that the transformations follow a uniform random selection, we estimate that the data can be augmented such that each image returns $360\cdot4^2\cdot3=17,280$ augmented images. In practice the computational cost of parsing the full data set $17,280$ times is extremely high. Consequently, to complete the training process within a reasonable amount of time, whilst maintaining the benefits of the data augmentation, the randomly transformed training data set is passed into the model $72$ times during each epoch.
This amounts to the model optimising on 72 random augmentations of the training data set at each epoch.
This means, that throughout training, the model will at most see $7,200$ separate data transformations of the estimated $17,280$ possible transformations. 
During evaluation and testing we use $360$ random data set transformations, as the computational cost is comparatively negligible without the optimisation stages.

The test sets are split from the original data sets, containing equivalent fractional populations of FRI and FRII sources, and are reserved for model evaluation. The remaining data samples are split randomly using an 80:20 training:validation split. The training set is used to optimise the network parameters, and the validation set is used for early stopping, i.e. the model that has the lowest validation loss throughout training is saved as the final model. Augmentation is applied after this split, ensuring that none of the unique sources from the test set appear within the training and/or validations sets.

Table~\ref{tab:dataset_summary} shows how the data sets are split for training, validation and testing, and the Total$^{\ast}$ column lists the augmented number of each data subset. This is done to provide an understanding of the scale of the relatively modest augmentation. It is important to note, that the value of data augmentation is not a replacement for the inclusion of additional unique sources, or larger data sets, but rather is a supplementary process which allows data invariances to be exploited and reduce biases in the generalised model. 
\begin{table*}
    \centering
    \begin{tabular}{|c|c l | c c c c c|}
        \hline
        Data Set & \multicolumn{2}{c|}{Label} & Total$^{\ast}$ & Total & Train & Validation & Test \\ [0.5ex]
        \hline\hline
        \multirow{3}{*}{FR-DEEP}
        & \multicolumn{2}{c|}{FRI}   & $19,008$ & 264 & 193.6 & 48.4 & 22 \\ [0.5ex]
        & \multicolumn{2}{c|}{FRII}  & $24,192$ & 336 & 246.4 & 61.6 & 28 \\ [0.5ex]
        & \multicolumn{2}{c|}{Total} & $43,200$ & 600 & 440   & 110  & 50 \\ [0.5ex] \hline\hline
        \multirow{6}{*}{MiraBest}
        & \multirow{2}{*}{FRI}  & Certain           & $28,584$  & 397  & 278 & 70  & 49 \\
        &                       & Uncertain         & $13,968$  & 194  & 135 & 34  & 25 \\[1.5ex]
        & \multirow{2}{*}{FRII} & Certain           & $31,392$  & 436  & 305 & 76  & 55 \\
        &                       & Uncertain         & $14,040$  & 195  & 137 & 34  & 24 \\ [1.5ex]
        & \multicolumn{2}{c|}{Total (excl. hybrid)} & $87,984$ & $1,222$  & 855 & 213 & 153 \\
        \hline
    \end{tabular}
    \caption{Summary of the respective subsets of FR-DEEP and MiraBest, as implemented in this work. The Total* column refers to the totals augmented by $72$ random transformations (the smallest number of random transformations applied). The training and validation data sets are split at random, and thus the values in the training and validation columns, which refer to specific (sub) classes, are approximate expectations rather than absolute values.}
    \label{tab:dataset_summary}
\end{table*}

\section{Model Performance}
\label{sec:performance}

To determine a baseline for our models on the MiraBest data set, we compare the performance of the \textit{AG-CNN} architecture trained on various data sets to the \textit{Classic CNN} model trained and evaluated in \cite{Tang2019a}. Table \ref{tab:Attention_vs_Classic} lists the results of the evaluation of the various models.
\begin{table*}
    \centering
    \begin{tabular}{|c|c c  c c  c c  c c|}
        \hline
        Network & \multicolumn{2}{c}{Classic CNN} & \multicolumn{2}{c}{AG-CNN} & \multicolumn{2}{c}{AG-CNN} & \multicolumn{2}{c|}{AG-CNN} \\
        Data Set  & \multicolumn{2}{c}{FR-DEEP-F} & \multicolumn{2}{c}{FR-DEEP-F} & \multicolumn{2}{c}{MiraBest*} & \multicolumn{2}{c|}{MiraBest} \\
        Class     & FRI          & FRII          & FRI    & FRII   & FRI    & FRII   & FRI    & FRII \\ [0.5ex] \hline\hline
        F1 Score  & $0.90\pm0.03$& $0.88\pm0.06$ & $0.87$ & $0.90$ & $0.91$ & $0.92$ & $0.82$ & $0.86$ \\
        Precision & $0.95\pm0.02$& $0.83\pm0.04$ & $0.87$ & $0.90$ & $0.89$ & $0.89$ & $0.91$ & $0.80$ \\
        Recall    & $0.85\pm0.02$& $0.94\pm0.04$ & $0.87$ & $0.90$ & $0.95$ & $0.94$ & $0.75$ & $0.93$ \\
        Accuracy & \multicolumn{2}{c}{$89\pm 1\,\%$} & \multicolumn{2}{c}{$88\,\%$} & \multicolumn{2}{c}{$92\,\%$} & \multicolumn{2}{c|}{$84\,\%$} \\
        AUC           & \multicolumn{2}{c}{$0.94$}      & \multicolumn{2}{c}{$0.89$} & \multicolumn{2}{c}{$0.96$} & \multicolumn{2}{c|}{$0.92$} \\
        \hline
    \end{tabular}
    \caption{A comparison of a classical CNN classifier (from \citealt{Tang2019a}) to an implemented attention gated network (with range normalisation, aggregated using the fine tuned method and 3 attention gates) trained and tested on the data set listed. \textit{MiraBest*} refers to the sub-set of MiraBest sources labelled as certain (see Section \ref{sec:datasets}). The evaluation metrics are clarified in Appendix~\ref{app:EvaluationMetrics}.}
    \label{tab:Attention_vs_Classic}
\end{table*}

From this table it can be seen that the AG-CNN architecture performs similarly to the Classic CNN on the FR-DEEP data set, and in the case where it is trained on the MiraBest$^{\ast}$ data set the resulting model shows an improved performance. These observations are notable as the AG-CNN uses fewer than half as many parameters in comparison to the Classic CNN: 101\,k compared with 250\,k parameters, see Table~\ref{tab:RadGalNet}. The relative performance loss seen for the model trained on the full MiraBest data set is expected, as the MiraBest-Uncertain subset of sources are more difficult to classify confidently for experts, and thus will cause larger error rates in the respective model, both because they contain edge cases which are more difficult to classify and because the labels of the sources themselves may not be fully correct.

\begin{figure}
    \centering
    \includegraphics[width=0.5\textwidth]{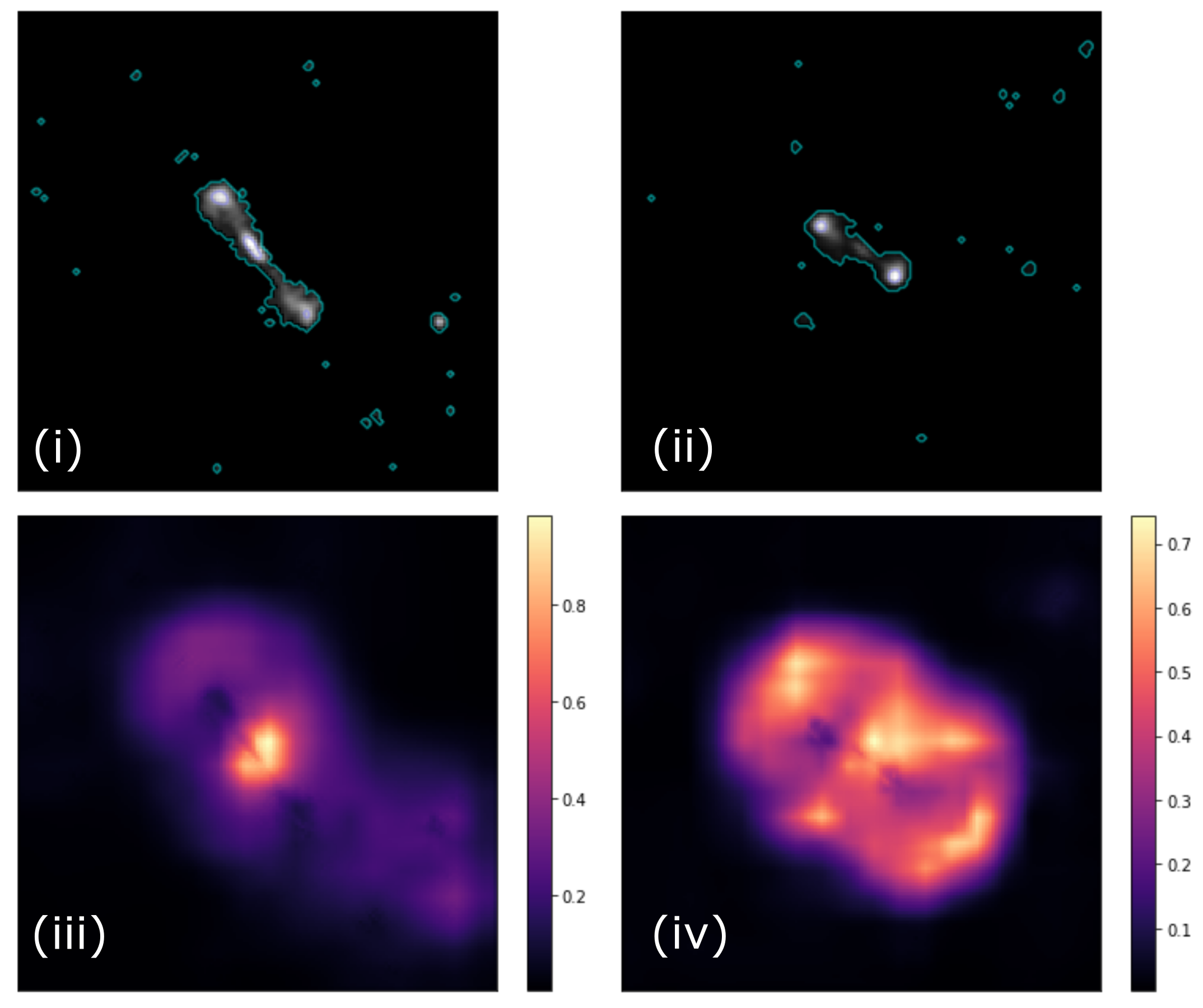}
    \caption{Example FRI and FRII sources selected for their respective clear classification with averaged attention maps derived from our model. (i) FRI example source: SDSS ID 2114-53848-625. (ii) FRII example source: SDSS ID 2266-53679-612. (iii) and (iv) are the averaged attention maps for the FRI and FRII sources respectively.}
    \label{fig:Example_Sources}
\end{figure}

Unlike traditional CNN models, the attention maps from the AG-CNN can also be visualised to understand how the model scales the features it extracts to make a final classification. To illustrate this we use two example sources from the MiraBest data set. As an example of the FRI class we use SDSS\,ID\,2114-53848-625  (J\,13h54m33.0s~+28$^{\circ}$14'36'') and as an example of the FRII class we use SDSS\,ID\,2266-53679-612 (J\,08h04m04.5s~+15$^{\circ}$33'35''). 
These sources and their combined attention maps are shown in Figure~\ref{fig:Example_Sources}. In this figure it can be seen that both attention maps attend the region immediately around the source of interest, with the FRI example also attending the immediate vicinity of a secondary source to the bottom right of the image. While the FRI attention map primarily attends the central source, the FRII attention map primarily attends the immediate surroundings of the source, which is where the features relating to the lobes are expected to be found. Both images attend the region around the source more than the source itself, as shown by the dark regions in the shape of the sources extending through the middle of each attended region. We note the difference in scale between the averaged maps, with the FRI source peaking at 0.98 and the peak of the FRII region peaking at 0.74. This is a consequence of the range normalisation used in the model and indicates that the peaks of the three individual attention maps align more closely in the FRI case, and less well in the FRII example.

\subsection{Normalisation and Aggregation}
\label{sec:normalisation_effects}

To investigate how the selection of normalisation function and aggregation function affect the model and its performance, we train multiple three gated attention models on the MiraBest data set. Table~\ref{tab:normalisation} shows how the model performs across each of the normalisations using the fine tuned aggregation method and Table~\ref{tab:normalisations_aggregations_averaged} shows how the model performs for each normalisation averaged across all aggregation methods and for each aggregation method, averaged across normalisation methods. These averaged evaluations show that the concatenation method has a slight performance gain over the other aggregation methods, and that softmax or sigmoid should be the primary choices for normalisation based on performance alone. However, whilst a model based purely on its performance may be of interest, for the attention mechanism to be maximally effective the interpretation of the attention maps themselves should also be taken into consideration.
\setlength{\tabcolsep}{3pt}
\begin{table}
    \centering
    \begin{tabular}{|c|c c  c c  c c  c c|}
        \hline
        Normalisation & \multicolumn{2}{c}{Range Norm.} & \multicolumn{2}{c}{Standardisation} & \multicolumn{2}{c}{Sigmoid} & \multicolumn{2}{c|}{Softmax} \\
        Class     & FRI   & FRII   & FRI    & FRII   & FRI    & FRII   & FRI    & FRII \\ [0.5ex] \hline\hline
        F1 Score  & $0.82$& $0.86$ & $0.64$ & $0.67$ & $0.85$ & $0.87$ & $0.86$ & $0.88$ \\
        Precision & $0.91$& $0.80$ & $0.65$ & $0.66$ & $0.90$ & $0.84$ & $0.92$ & $0.84$ \\
        Recall    & $0.76$& $0.93$ & $0.63$ & $0.68$ & $0.81$ & $0.91$ & $0.81$ & $0.93$ \\
        Avg. Accuracy & \multicolumn{2}{c}{$84\,\%$} & \multicolumn{2}{c}{$66\,\%$} & \multicolumn{2}{c}{$87\,\%$} & \multicolumn{2}{c|}{$87\,\%$} \\
        AUC           & \multicolumn{2}{c}{$0.92$}      & \multicolumn{2}{c}{$0.66$} & \multicolumn{2}{c}{$0.92$} & \multicolumn{2}{c|}{$0.93$} \\
        \hline
    \end{tabular}
    \caption{Evaluations of attention model with each of the four normalisations (fine tuned aggregation and three attention gates).}
    \label{tab:normalisation}
\end{table}
\setlength{\tabcolsep}{3pt}
\begin{table}
    \footnotesize
    \centering
    \begin{tabular}{|l|c c  c c  c c  c c|}
        \hline
        Norm. & \multicolumn{2}{c}{Range Norm.} & \multicolumn{2}{c}{Standardisation} & \multicolumn{2}{c}{Sigmoid} & \multicolumn{2}{c|}{Softmax} \\
        Class     & FRI   & FRII   & FRI    & FRII   & FRI    & FRII   & FRI    & FRII \\ [0.5ex] \hline\hline
        F1 Score  & $0.81$& $0.83$ & $0.69$ & $0.71$ & $0.83$ & $0.84$ & $0.85$ & $0.87$ \\
        Precision & $0.84$& $0.80$ & $0.72$ & $0.70$ & $0.85$ & $0.83$ & $0.89$ & $0.84$ \\
        Recall    & $0.77$& $0.87$ & $0.67$ & $0.73$ & $0.81$ & $0.86$ & $0.81$ & $0.91$ \\
        Accuracy & \multicolumn{2}{c}{$82\,\%$} & \multicolumn{2}{c}{$70\,\%$} & \multicolumn{2}{c}{$84\,\%$} & \multicolumn{2}{c|}{$86\,\%$} \\
        AUC           & \multicolumn{2}{c}{$0.87$} & \multicolumn{2}{c}{$0.71$} & \multicolumn{2}{c}{$0.85$} & \multicolumn{2}{c|}{$0.92$} \\
        \hline\hline
        Agg. & \multicolumn{2}{c}{Mean} & \multicolumn{2}{c}{Concatenation} & \multicolumn{2}{c}{Deep Supervised} & \multicolumn{2}{c|}{Fine Tuned} \\
        Class     & FRI   & FRII   & FRI    & FRII   & FRI    & FRII   & FRI    & FRII \\ [0.5ex] \hline\hline
        F1 Score  & $0.78$& $0.82$ & $0.82$ & $0.84$ & $0.79$ & $0.78$ & $0.79$ & $0.82$ \\
        Precision & $0.85$& $0.77$ & $0.86$ & $0.81$ & $0.77$ & $0.80$ & $0.83$ & $0.79$ \\
        Recall    & $0.72$& $0.88$ & $0.78$ & $0.88$ & $0.81$ & $0.75$ & $0.76$ & $0.85$ \\
        Accuracy & \multicolumn{2}{c}{$80\,\%$} & \multicolumn{2}{c}{$83\,\%$} & \multicolumn{2}{c}{$78\,\%$} & \multicolumn{2}{c|}{$81\,\%$} \\
        AUC           & \multicolumn{2}{c}{$0.83$} & \multicolumn{2}{c}{$0.86$} & \multicolumn{2}{c}{$0.82$} & \multicolumn{2}{c|}{$0.85$} \\
        \hline
    \end{tabular}
    \caption{Evaluations for each normalisation and aggregation. The sets of models are averaged across aggregation to produce individual scores for the normalisations, and grouped by normalisation to produce individual scores for the aggregations.}
    \label{tab:normalisations_aggregations_averaged}
\end{table}

\begin{figure}
    \centering
    \centerline{(i) FRI}
    \includegraphics[width=\linewidth]{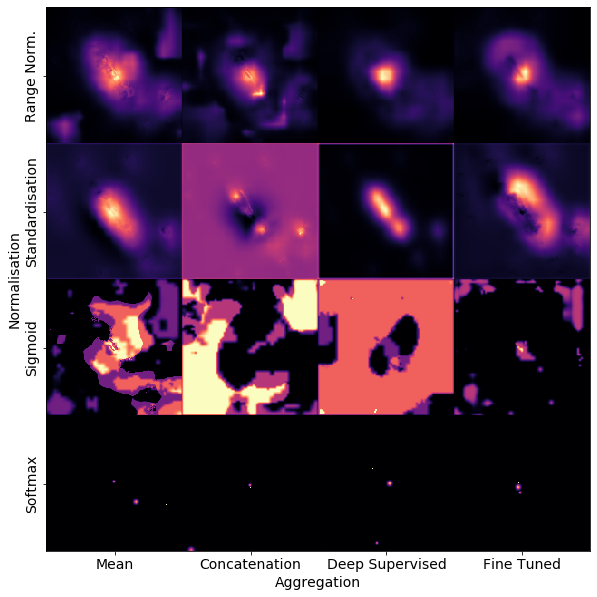}
    \centerline{(ii) FRII}
    \includegraphics[width=\linewidth]{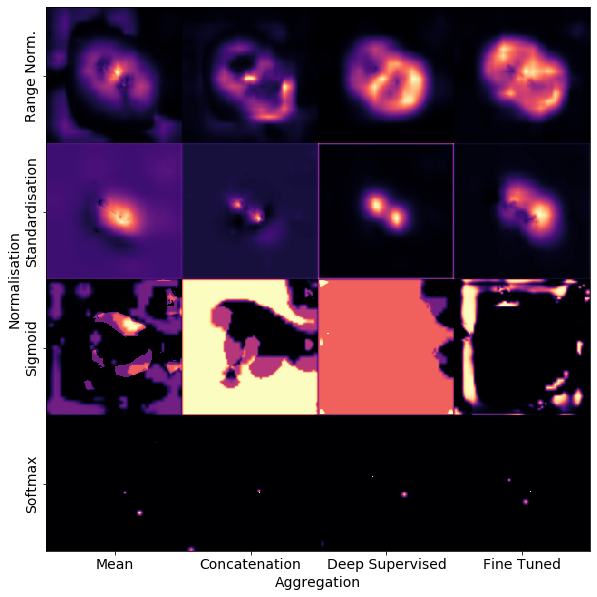}
    \caption{Array of averaged attention maps for (i) the FRI example source, and (ii) the FRII example source. The models each have three attention gates and differ by selection of normalisation and aggregation methods.}
    \label{fig:Attention_array_FRI}
\end{figure}

We present in Figure~\ref{fig:Attention_array_FRI} the attention maps produced from different normalisation and aggregation modes made using the example FRI/II sources that were presented in Figure~\ref{fig:Example_Sources}. From this figure it is clear that the marginal improvements in performance seen for the softmax and sigmoid methods are gained at the expense of clarity and interpretability of the attention maps themselves. This result is consistent with the findings of \cite{Schlemper2019}, who moved away from the use of softmax as a normalisation method for attention gating due to the sparsity of the resulting attention maps. 

\subsection{Attention Gate Number}
\label{sec:no_AG_effects} 

To investigate the impact of including different numbers of attention gates, we consider models trained on the MiraBest data set using the range normalisation and fine tuned aggregation methods with a varying number of attention gates. Table~\ref{tab:No_AG_eval} displays the resulting evaluations of the respective models.
\setlength{\tabcolsep}{5pt}
\begin{table}
    \centering
    \footnotesize
    \begin{tabular}{|l|c c  c c  c c  c c|}
        \hline
         Gates & \multicolumn{2}{c}{0} & \multicolumn{2}{c}{1} & \multicolumn{2}{c}{2} & \multicolumn{2}{c|}{3} \\
        Class     &  FRI   &  FRII   &  FRI   & FRII   &  FRI   &  FRII   &  FRI  &  FRII   \\ [0.5ex] \hline\hline
        F1 Score  & $0.85$& $0.88$ & $0.84$ & $0.85$ & $0.86$ & $0.88$ & $0.83$ & $0.86$ \\
        Precision & $0.91$& $0.83$ & $0.86$ & $0.83$ & $0.91$ & $0.84$ & $0.91$ & $0.80$ \\
        Recall    & $0.80$& $0.93$ & $0.81$ & $0.88$ & $0.81$ & $0.93$ & $0.76$ & $0.93$ \\
        Accuracy & \multicolumn{2}{c}{$86\,\%$} & \multicolumn{2}{c}{$85\,\%$} & \multicolumn{2}{c}{$87\,\%$} & \multicolumn{2}{c|}{$85\,\%$} \\
        AUC           & \multicolumn{2}{c}{$0.93$} & \multicolumn{2}{c}{$0.90$} & \multicolumn{2}{c}{$0.94$} & \multicolumn{2}{c|}{$0.92$} \\
        \hline
    \end{tabular}
    \caption{Evaluations for models trained on MiraBest with varying number of attention gates (see Section \ref{sec:attention_implementation} for clarification). The model with 1 attention gate was trained with a learning rate of $10^{-5}$ instead of the otherwise used $5\cdot10^{-5}$.}
    \label{tab:No_AG_eval}
\end{table}

\begin{figure}
    \centering
    \includegraphics[width=\linewidth]{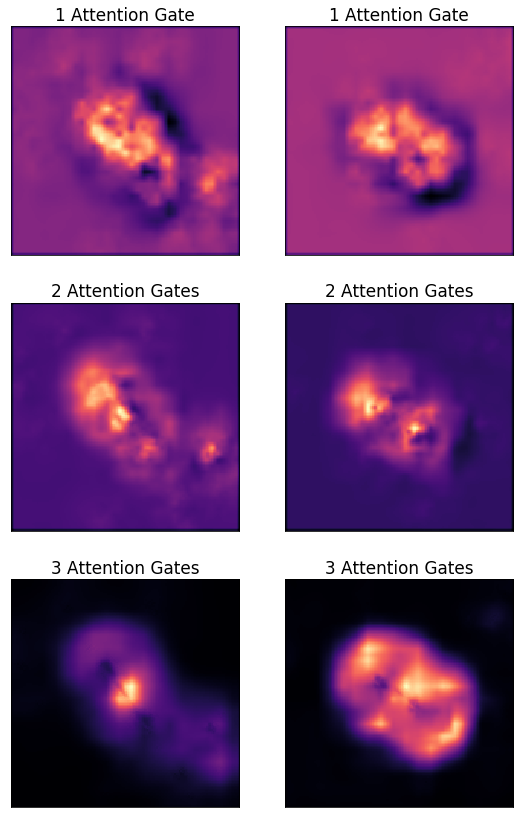} 
    \caption{The (averaged) attentions of our models implemented with varying number of attention gates for the FRI (left) and FRII (right) example sources, as introduced in Figure \ref{fig:Example_Sources}.}
    \label{fig:No_AG_Attentions}
\end{figure}

In this case, the highest performing model in terms of accuracy is the model with two attention gates. However, as with the investigation of aggregation and normalisation methods, the resulting attention maps also play a part in evaluating the value of a given model. Figure~\ref{fig:No_AG_Attentions} visualises the average attention maps when considering the example FRI and FRII sources. Here, one can see that the saliency achieved by the models with 1 and 2 attention gates is more dispersed. Although this does not hinder the model's performance, it is potentially more difficult for a human observer to interpret.

\cite{Schlemper2019} state that they empirically found a third attention gate did not provide additional value to the system, but \cite{Jetley2018} state that the third attention gate encourages the model to learn salient features earlier in the network, as these features are used to make a third of the classification since they used mean aggregation. We suggest that the selection of attention gates should be considered with reference to the specific data problem and that the ability of the user to relate to the resulting attention distributions should be considered a factor in this process.

\subsection{Attention as a Function of Epoch}
\label{sec:Attention_Epoch}

\begin{figure*}
    \centering
    \includegraphics[width=\linewidth]{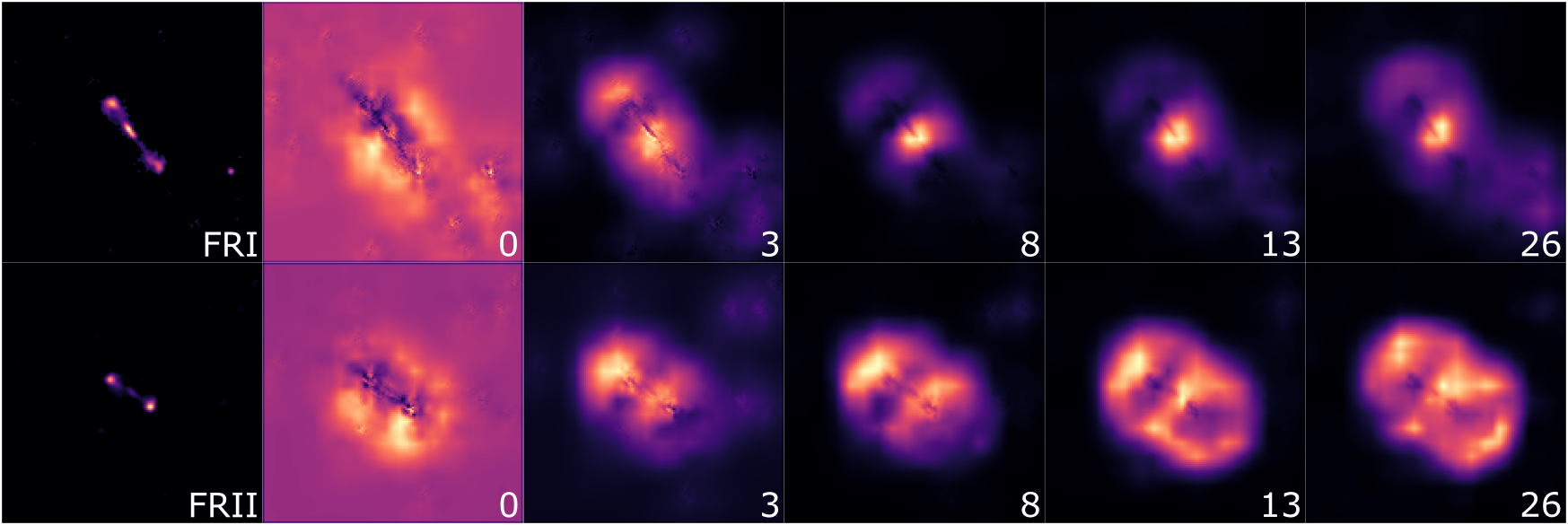}
    \caption{From left to right: Example sources from Figure~\ref{fig:Example_Sources} followed by their attention maps at the respective epoch number throughout the training process. The epochs are sampled uniformly across the epochs where the model's minimal validation loss was improved. Epoch 26 produced the model with the minimal validation loss, and thus the final model.}
    \label{fig:AMaps_by_Epoch}
\end{figure*}

Figure~\ref{fig:AMaps_by_Epoch} shows the two example sources and their attention maps at different points throughout the training process. As a function of epoch, the attention develops to become more specific to the regions discussed previously: the attention maps show no clear feature selection at epoch 0, but become more specialised throughout training. Note that in the FRI case, after the model learns to attend the central source (epoch 8) it shifts to attend the region around the source to a higher degree than it had previously. For the FRII case, the attended regions of epochs 3 and 8 are distributed asymmetrically around the source, with a larger focus on an offset region above the source's lobes. By epoch 26, the learned attention is much more symmetrical for this example.

\section{Aggregate Attention}
\label{sec:distr_attention}

While presenting the attention maps for example sources is helpful when considering the model in individual cases, we must also consider the population average of attention across the data set. We calculate these averaged images by considering the mean of a randomly transformed set of images: both the test set itself as well as the attention maps derived from the augmented test set. Each attention map in this section is created by taking the pixel mean across a set of attention maps generated by passing the augmented test set into the model 100 times. We note that the statistics of the attention distribution across the augmentations is not Gaussian and will be considered in more detail in future work. As such, the mean is used as a comparative measure rather than a true parameterisation.

\begin{figure}
    \centering
    \includegraphics[width=0.48\textwidth]{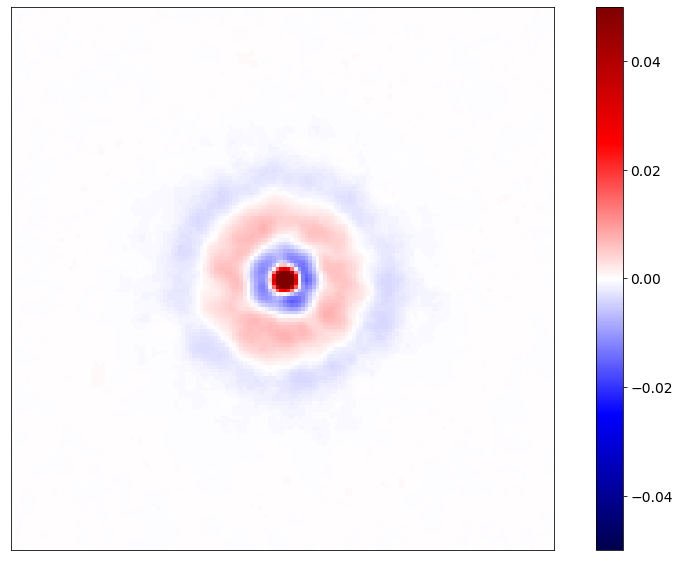}
    \caption{The difference between the mean pixel values of the 100 fold augmented FRI and FRII MiraBest test sets, $\langle\text{FRI}\rangle - \langle\text{FRII}\rangle$. To emphasise the structures in this difference image it is saturated at $\pm0.05$, with an original maximal difference of $0.1$ at the centre.}
    \label{fig:Distributions_Sources}
\end{figure}

For the input images themselves, Figure~\ref{fig:Distributions_Sources} presents the difference between the mean pixel intensities of the FRI and FRII classes. Although there is clear structure, the individual sources are far less well defined than this figure may suggest, and we therefore caution against over-interpretation. However, a broad interpretation suggests that the central pixels are dominated by FRI sources, as one would expect from the morphological definition of that population. A first ring forms due to the lobes of FRII samples. A second ring forms due to the extended jet emission of FRI sources, which are some of the brightest regions of the given source image. This effect is likely enhanced due to the min-max normalisation applied to each image. A third ring forms due to the peak emission of FRII sources often being the hot spots in the extended lobes, which tend to be much brighter than their jets.

\begin{figure}
    \centering
    \includegraphics[width=0.5\textwidth]{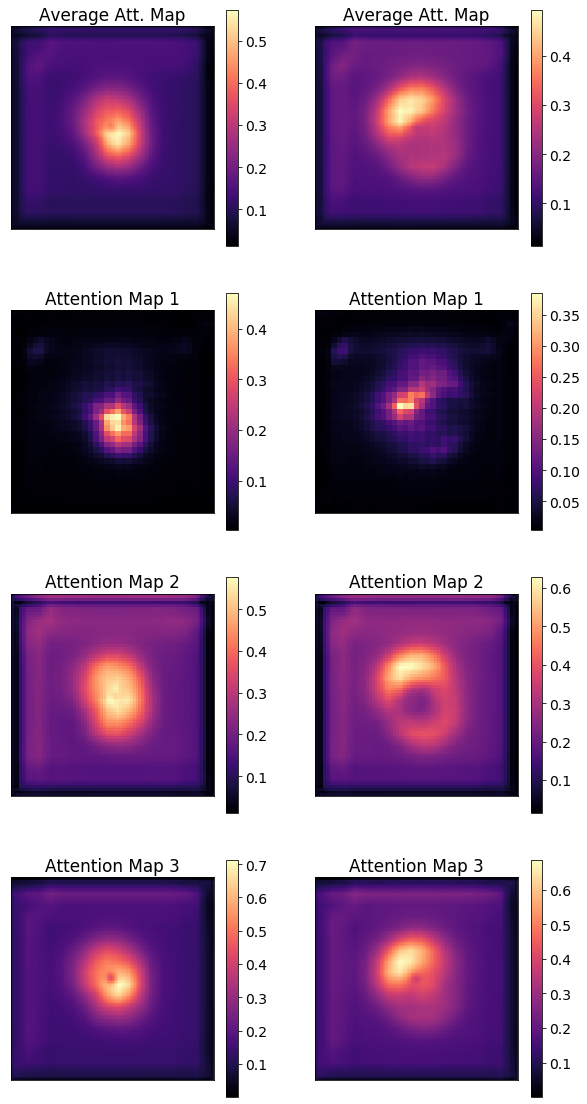}
    \caption{Mean pixel values of the average and individual attention maps of the test set after 100 fold augmentation. Left: FRI; Right: FRII.}
    \label{fig:Distr_AGates}
\end{figure}

Figure~\ref{fig:Distr_AGates} shows the difference in attention when separating the mean pixel intensities by class across the whole of the augmented MiraBest test set. The aggregate attention map shows a ring-like structure for FRII sources, which seems to stem primarily from attention gate 2. For the FRI sources, a centred Gaussian attention is more prevalent. As FRII sources are typically classified on their lobes and FRI sources are classified on the brightness of their central engines, this aligns well with how a human classifier would attend a data set on average.

\begin{figure*}
    \centering
    \includegraphics[width=0.95\textwidth]{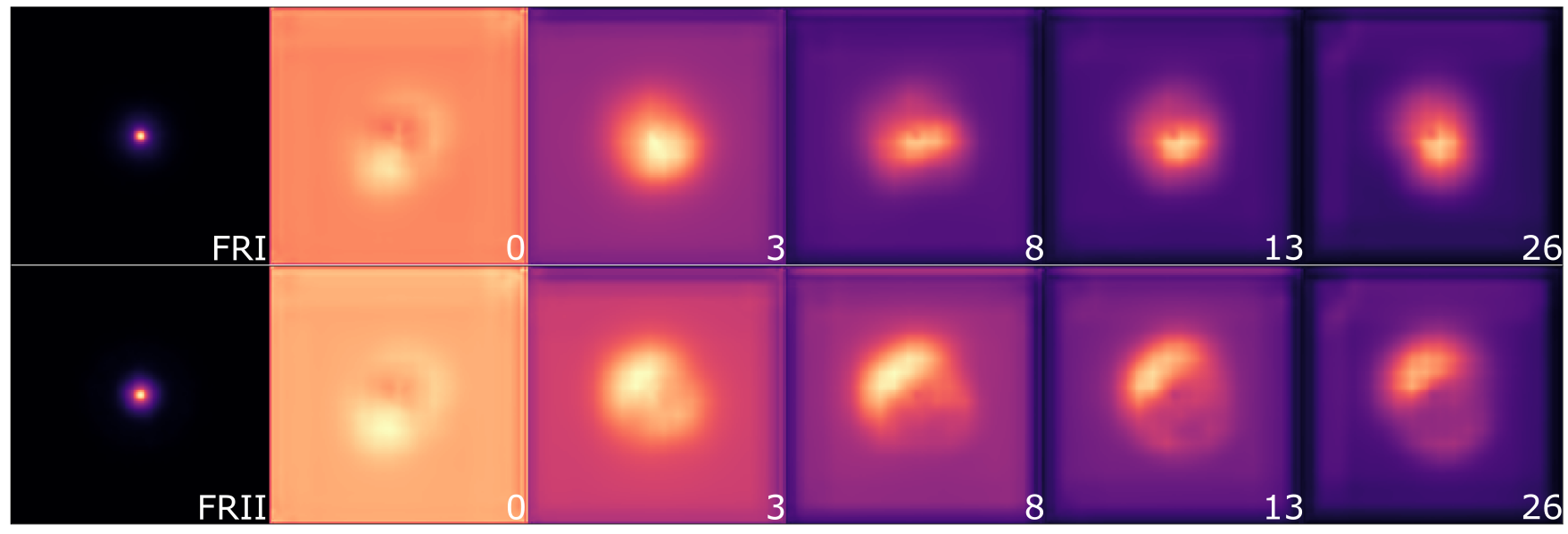}
    \caption{Mean pixel values of sources and attention maps throughout training at given epochs after 100 fold augmentation. See Figure \ref{fig:AMaps_by_Epoch} for the equivalent plot with the example sources.}
    \label{fig:distr_attention_epoch_binary}
\end{figure*}

By considering the aggregated attention across various subsets, a number of insights can be gained. Analogous to the discussion in previous section, Figure~\ref{fig:distr_attention_epoch_binary} shows how the aggregate attention develops throughout training. At epoch 0 the aggregate attention is highly similar across the two classes, but as the epochs progress the ring and Gaussian shapes develop in the aggregate attention maps of the respective classes, demonstrating how clearly the two classes are separated by the model's attention before any fully connected layers are applied.

\begin{figure}
    \centering
    \includegraphics[width=0.49\textwidth]{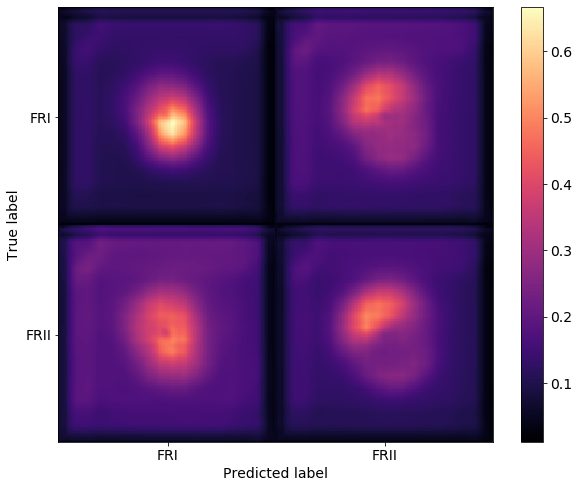}
    \caption{Distribution of attention maps according to the predicted and true labels of the respective sources.}
    \label{fig:distr_confusion_matrix}
\end{figure}

We also consider that it may be possible to see a difference in the aggregate attention maps for sources that are predicted correctly and those which the model predicts incorrectly. To visualise this, Figure~\ref{fig:distr_confusion_matrix} shows a `confusion matrix' of aggregated attention maps. It can be seen that the sources which were predicted incorrectly present less clearly defined attention with the largest difference being in the size of the central region and the direction of the offset of the bright spot in the attended areas which aligns with the predicted class.

\begin{figure}
    \centering
    \includegraphics[width=0.5\textwidth]{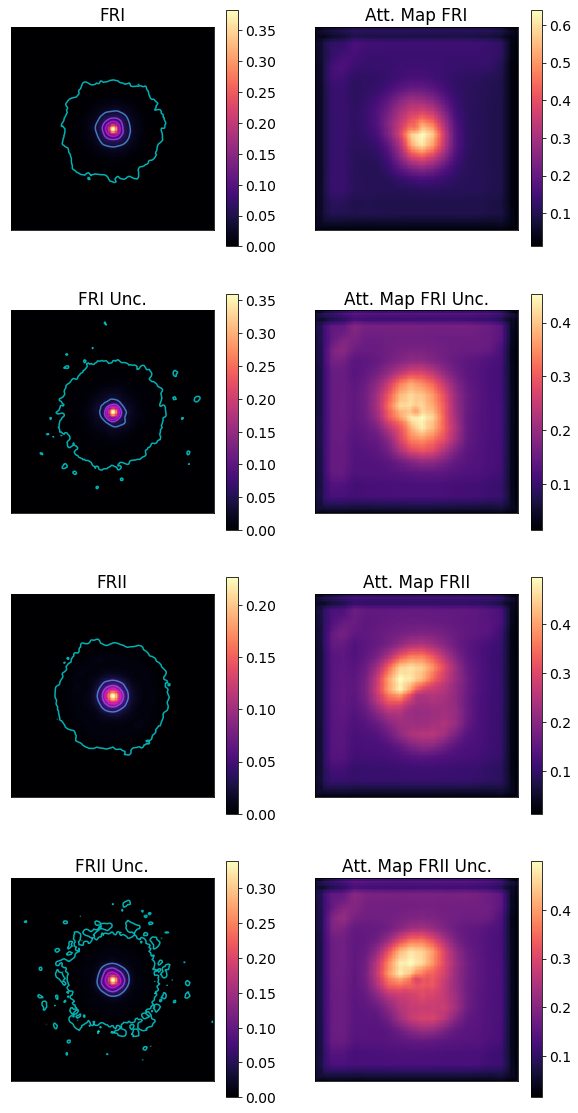}
    \caption{Mean source intensities and attention maps according to both class and certainty. With contours plotted alongside the mean source mean intensity to highlight class differences at 0.001, 0.005, 0.01, 0.05 and 0.1 respectively.}
    \label{fig:distr_subsets}
\end{figure}

Similarly, the uncertainty in the data set becomes clear when considering the mean pixel intensities of the MiraBest-Certain and MiraBest-Uncertain attention maps and test data. Figure~\ref{fig:distr_subsets} shows these maps and it can be seen that the attention maps are significantly less distinct for the MiraBest-Uncertain sources. The mean pixel intensities themselves can be separated by the peaks of the respective maps, which do not align with those of the confident FRI and FRII sources. The FRII peak is expected to be higher, as FRII sources with brighter cores are expected to be `uncertain'. The uncertain FRI source mean intensity map has a higher peak than the certain FRI map. This may be due to the inherent classification bias, where fringe source cases with bright centres are more likely to be classified as FRI even though they are inherently uncertain.

\begin{figure}
    \centering
    \centerline{(i) FRI}
    \includegraphics[width=\linewidth]{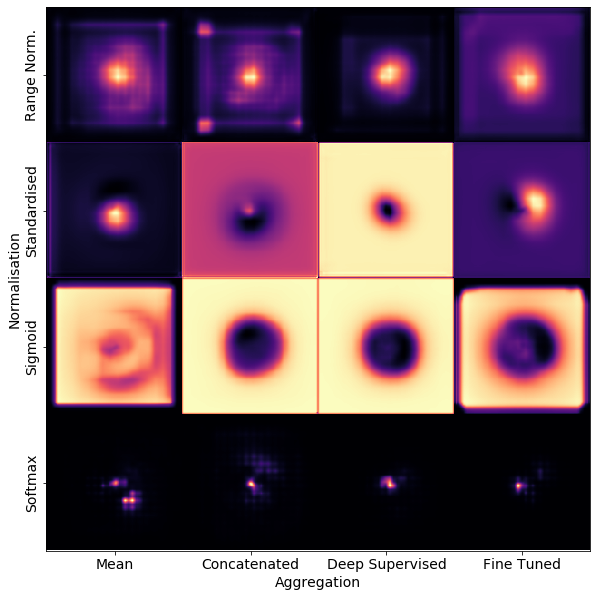}
    \centerline{(ii) FRII}
    \includegraphics[width=\linewidth]{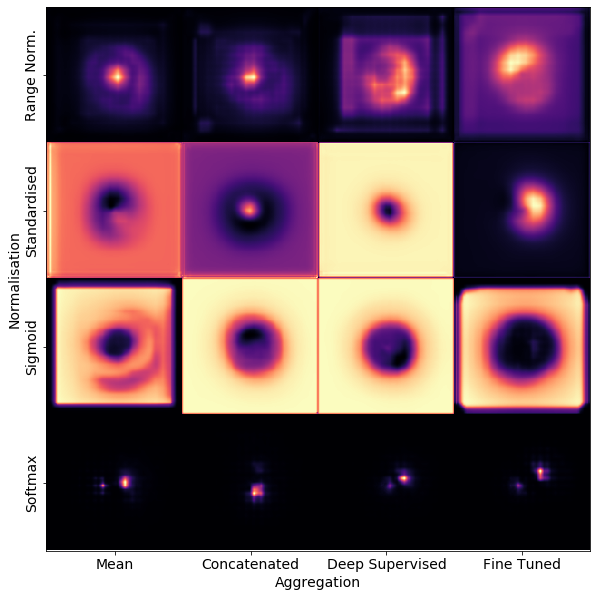}
    \caption{Aggregated attention maps of models implemented with various aggregations and normalisations on the (i) FRI and (ii) FRII MiraBest data sets.}
    \label{fig:distr_AggNorm}
\end{figure}

Finally, the aggregate attention maps of the 16 models trained with each permutation of aggregation and normalisation, introduced in Section~\ref{sec:normalisation_effects}, are shown in Figure~\ref{fig:distr_AggNorm}. In the case of the individual exemplar sources, see Figure~\ref{fig:Attention_array_FRI}, the sigmoid models seem to be attending regions that an observer cannot clearly recognise; however in the case of the aggregate attention maps it is clear that they tend to highlight zero space around the source distributions, with little differences between the respective FRI and FRII distributions. 

Similarly to the case of individual sources, the softmax normalised aggregate attention map is not helpful. Beyond the softmax models, all of the models presented demonstrate aggregate attention focused either on the central region of the image (range normalised models and some standardised models) or attend the zero spaces around the mean source pixel intensities (sigmoid models and some standardised models). Although knowing that the sigmoid models are attending regions representative of the mean source pixel intensity maps builds confidence in the models, their individual attention maps are not helpful for individual source analysis, see Figure~\ref{fig:Attention_array_FRI}.

The value of these mean intensity and aggregated attention maps not only lies in the analysis of the model and understanding the difference in how it attends various subsets, but also in the assistance they provide when developing new models, as they can be used to analyse the data set distributions and evaluate a model's ability to generalise to unseen testing data.

\section{Individual Sources}
\label{sec:individual_sources}

Using this work's primary model, which implements range normalisation, fine tuned aggregation, and three attention gates trained on the MiraBest data set, we highlight some specific sources of potential interest. To do this, we evaluate the test set across 1,000 random augmentations, as described in Section~\ref{sec:data_augmentation}, and calculate the false classification rate for each test source. 

\begin{figure*}
    \centering
    \includegraphics[width=\linewidth]{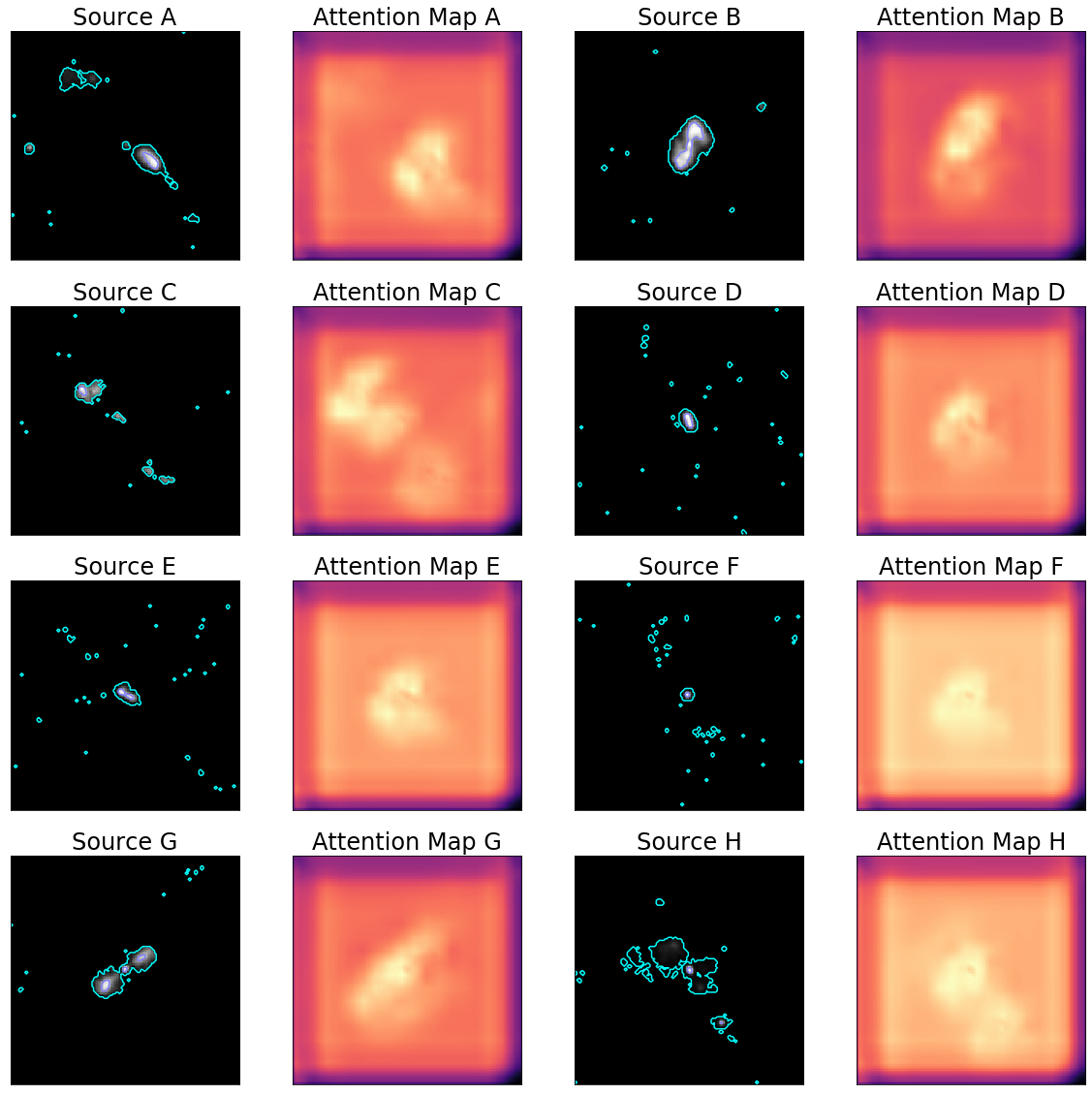}
    \caption{Sources which have a false classification rate above 95\,\%. The pre-processed source image, with a contour highlighting the sigma clipped zero space, and their respective attention maps. Sources are grouped by class label and certainty. Details for each source can be found in Table~\ref{tab:FalselyClassifiedDetails}.}
    \label{fig:FalselyClassifiedSources}
\end{figure*}

By separating out sources that are classified incorrectly for more than 95\,\% of the augmentations, we highlight the objects which fundamentally do not align with the model's learned classification. Details of these sources are given in Table~\ref{tab:FalselyClassifiedDetails} and  Figure~\ref{fig:FalselyClassifiedSources} presents their images and respective attention maps. 

Sources A and B are the only certain FRI sources mis-classified as FRII sources at a higher rate than 95\,\% by the model. The attention map of source A shows that the model is attending the features of the bright source near the centre of the image; however, it is unclear whether this bright source is aligned with the host galaxy for this source. If it is, then this source seems to be miss-labelled, as the only visible jet would be edge brightened. 
If the bright source itself is the central engine, then the source would be classified as an FRI. The model's training procedure could be adapted to allow for significantly off-centre sources by augmenting the data such that the central engine of each source could be anywhere on the image. 
This is an example of how attention maps can help inform the training of the model and the selection of the training data.

The class of source B is not immediately clear from a simple visual inspection. 
The jets do not have individual bright spots, but rather have bright regions which makes the classification less clear. It may be that the training data simply did not contain any cases of FRI sources where the jets showed bright extended regions (instead of hot spots) which extended into the largest part of the extent of the source. This could be remedied by training with more data, which would hopefully `fill in the gaps' which the model still has.

Sources C-G are classed as uncertain FRI sources. Sources D, E and F are not well resolved and it is difficult to specify which class they belong to (assuming more distant flux in the image is not from the central radio galaxy itself). For source C, the model may in fact be classifying a single lobe of the radio galaxy, as indicated by the enhancement in the attention map to the top left of the image. The model is not aware that the source has been extracted to be at the centre of the image and may thus classify the lobe as an FRII radio galaxy.

Source G is not a bent source, nor is it unresolved or strange in any other way. It does however seem difficult (on first viewing) to see whether or not it is an FRI or FRII, as the distance ratio is not clear on first viewing, but would require explicit measurement.

Source H is labelled as an uncertain FRII within the MiraBest data set, and is misclassified with a rate of over 95\,\% as an FRI by the model. The model tends to over-classify sources as FRII sources due to the slight imbalance in the training data, which makes this object a notable exception as the only FRII source mis-classified as FRI. A central bright-spot is clearly visible, but unlike the exemplar FRII source displayed in Figure~\ref{fig:Example_Sources}, source H shows no clear bright-spots in its wide-angled lobes. Mis-classification in this case may arise from the unusually bright central source, which itself may be a chance alignment with another galaxy along the line of sight. Alternatively it may be due to confusion with the secondary source at the bottom right of the image, which is associated with an enhanced area of attention.

Some of these sources may require more intentional efforts to be well understood and we note that outlining what could have caused the model to classify these sources incorrectly is not an attempt at physical analysis. This evaluation of the outlying data samples is meant to serve in two aspects. Firstly it is meant to highlight sources which may be of interest for improving the performance of the model. Secondly, this evaluation is meant to demonstrate the value of attention and machine learning for extracting individually abnormal sources, even from a seemingly simple data set, in the same way that \cite{Sasmal2020} manually selected and highlighted abnormal sources from the LOTSS data release 1. Future applications of this technique might also consider including data sets at other wavebands to clarify these aspects, in the same way that \citet{Wu2019} combined FIRST and WISE data for the CLARAN classifier.
\begin{table*}
    \centering
    \begin{tabular}{|c c c c c r c|}
        \hline
         Source & SDSS ID & RA & Dec & Redshift & Extent & MiraBest Class \\
        \hline\hline
        A & 1737-53055-197 & 07h48m18.9s & +45$^\circ$44$^\prime$46$^{\prime\prime}$ & 0.1850 & 228.03$^{\prime\prime}$ & Certain FRI \\
        B & 0875-52354-521 & 10h40m22.5s & +50$^\circ$56$^\prime$25$^{\prime\prime}$ & 0.1539 & 44.61$^{\prime\prime}$ & Certain FRI \\
        C & 1618-53116-159 & 11h29m54.5s & +06$^\circ$53$^\prime$12$^{\prime\prime}$ & 0.1162 & 120.67$^{\prime\prime}$ & Uncertain FRI \\
        D & 2598-54232-336 & 12h24m46.7s & +18$^\circ$25$^\prime$32$^{\prime\prime}$ & 0.1690 & 14.13$^{\prime\prime}$ & Uncertain FRI \\
        E & 2239-53726-557 & 12h51m57.1s & +30$^\circ$09$^\prime$26$^{\prime\prime}$ & 0.2235 & 11.94$^{\prime\prime}$ & Uncertain FRI \\
        F & 2510-53877-594 & 11h50m03.7s & +25$^\circ$39$^\prime$26$^{\prime\prime}$ & 0.1561 & 128.69$^{\prime\prime}$ & Uncertain FRI \\
        G & 1690-53475-090 & 16h49m24.0s & +26$^\circ$35$^\prime$03$^{\prime\prime}$ & 0.0545 & 68.49$^{\prime\prime}$ & Uncertain FRI \\
        H & 2203-53915-518 & 16h30m16.6s & +14$^\circ$35$^\prime$11$^{\prime\prime}$ & 0.2790 & 106.72$^{\prime\prime}$ & Uncertain FRII \\ \hline
    \end{tabular}
    \caption{\label{tab:FalselyClassifiedDetails} Details for the sources with a misclassification rate above 95\,\%, as depicted in Figure~\ref{fig:FalselyClassifiedSources}. Separated by dashes, the SDSS IDs are composed of the source's plate ID, Julian date and Fibre ID.}
\end{table*}

\section{Conclusions}
\label{sec:conclusion}

In this work we introduce attention as a state of the art mechanism for classification of radio galaxies using convolutional neural networks. We present an attention-based model that performs on par with previous classifiers while using over 50\% fewer parameters than the next smallest classic CNN application in this field. Furthermore, the AG-CNN presented in this work provides the additional benefit of visualising the salient regions used by the model in each case to make individual classifications. 

The primary model in this work was implemented using range normalisation, fine tuned aggregation and three attention gates. The model primarily attends the central engine for FRI sources, and increasingly attends the outer regions (i.\,e. lobes) of FRII sources. We observe that the salient regions identified by the attention gated model align well with the regions an expert human classifier would attend to make equivalent classifications. This includes both the central engine of the respective sources, the hot spots and the lobes of the source itself. This is also shown to be a learned trait of the model, by explicitly considering how the model's attention develops throughout training.

We also investigate how the selection of normalisation and aggregation methods used in attention-gating affect the output of individual models, using both quantitative evaluation metrics and the resulting attention maps to determine how employable each resulting model may be. Although the selection of such parameters minimally affects the model's performance, it can adversely affect individual models with regard to the interpretability of their respective attention maps. By selecting a model which aligns with how astronomers classify radio sources, the user can then employ the model to investigate what features the model is using to make classifications, and thus investigate how future models may be improved. We find that the softmax normalisation and concatenated aggregation methods provide the best model performance, but suggest that the range normalisation and fine-tuned aggregation methods provide the user with significantly improved attention maps at the cost of a minimal difference in performance. Similarly we find that the inclusion of a third attention gate does not contribute significantly to model performance but does aid in the interpretability of the resulting attention maps. 

We evaluate the average performance of models across the entire data set and use the example of individual sources in order to illustrate how attention maps can help the user engage in classification. By considering the aggregate attention maps across the augmented test set we note significant differences between various subsets of the data including the fundamental class division between FRI and FRII as well as that between correctly and incorrectly classified sources within each class. 

Finally, we present a method through which deep learning models can highlight individual sources for further study by extracting test sources that were found to be significantly misaligned with the predictions of the trained model. In these cases the availability of the attention maps can be used to examine the cause of the mis-classification in each case, as well as to understand of how complex the data might be, and how difficult even binary classification can become in certain cases. When applied to labelled data, this approach might also be used to select potential targets for future research by identifying sources that the model seems to consider as significant outliers.

\section*{Acknowledgements}

The authors would like to thank the anonymous reviewer for useful feedback that improved the contents of this paper.
MB gratefully acknowledges support from the University of Manchester STFC CDT in Data Intensive Science, grant number ST/P006795/1. AMS gratefully acknowledges support from an Alan Turing Institute AI Fellowship EP/V030302/1. FP gratefully acknowledges support from STFC and IBM through the iCASE studentship ST/P006795/1. DJB gratefully acknowledges support from STFC and the Newton Fund through the DARA Big Data program under grant ST/R001898/1.

\section*{Data Availability}

All code and trained models from this work are publicly available on Github at the following  address: \url{github.com/mb010/AstroAttention}. MiraBest is available on Zenodo: \url{10.5281/zenodo.4288837}. FR-DEEP is available on Zenodo: \url{10.5281/zenodo.4255826}.



\bibliographystyle{mnras}
\bibliography{paper_refs} 




\appendix

\section{Classification Evaluation Metrics}\label{app:EvaluationMetrics}
Classifications made by any model will be true (T) or false (F) as well as positive (P) or negative (N) for a given class. Knowing this, models can be evaluated by how many predicitons fall into each of the four subsets: TP, TN, FP and FN.
\citet{Gron2017_MLBook} contains a helpful overview of the metrics introduced here.

\subsubsection{Accuracy}
$\text{Accuracy}\in[0,1]$ is the ratio between correct predictions and all predictions. For data sets where the class sizes are not equal, accuracy should be calculated on a per class basis:
\begin{equation}
    \text{Accuracy} = \frac{\text{TP} + \text{TN}} {\text{TP} + \text{TN} + \text{FP} + \text{FN}}.
\end{equation}

\subsubsection{Precision}
$\text{precision}\in[0,1]$ is the ratio of positive classifications of all the positive classifications made.:
\begin{equation}
    \text{Precision} = \frac{\text{TP}}{\text{TP}+\text{FP}}.
\end{equation}

\subsubsection{Recall}
$\text{Recall}\in[0,1]$ is the proportion of positive samples which are classified positively. Recall is equivalent to class specific accuracies in the binary case:
\begin{equation}
    \text{Recall} = \frac{\text{TP}}{\text{TP}+\text{FN}}.
\end{equation}

\subsubsection{F1 Score}
$\text{F}_1\in[0,1]$ is the harmonic mean of precision and recall, averages their respective reciprocals. This is done to ensure that if either precision or recall is low, the F1 score suffers.
\begin{equation}
    \text{F}_1 = \frac{2}{\text{Precision}^{-1} + \text{Recall}^{-1}} = \frac{2\text{TP}}{2\text{TP} + \text{FP} + \text{FN}}
\end{equation}

\subsubsection{ROC}
\begin{figure}
    \centering
    \includegraphics[width=0.5\textwidth]{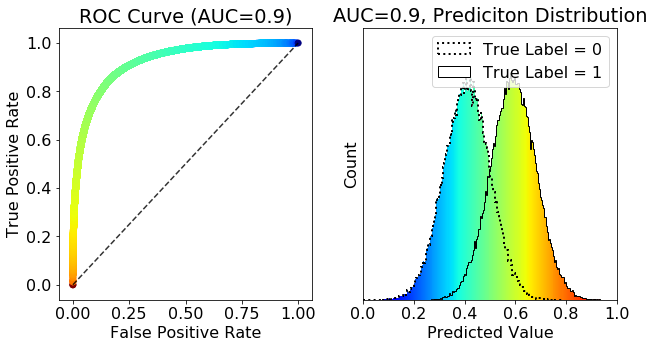}
    \caption{ROC Curve and the distributions which created it. Each distribution is a normal distribution of predicted class values in $[0,1]$. The threshold of a given point on the ROC curve is equivalent to the vertical color of the distributions. The largest class separation occurs at $\text{Predicted Value} \approx 0.5$ (green), which provides the optimal trade-off between TPR and FRP in this example case.}
    \label{fig:ROCCurve_Thresholds}
\end{figure}

The Receiver Operator Characteristic (ROC) curve is the name given to the curve which results from considering true positive rates (equivalent to recall) and the false positive rates:
\begin{equation}
\text{FPR} = \frac{\text{FP}}{\text{TN}+\text{FP}},
\end{equation}
at various thresholds. Thresholds are the value at which the two classes are separated. This is often $0$ in the case of predictions in the range of $[-1,1]$, or $0.5$ in the case of $[0,1]$. The threshold values for a example prediction distribution are visualised in Figure~\ref{fig:ROCCurve_Thresholds}. Given a binary classification, the ROC curve plots the recall of one class against the recall of the other.

\subsubsection{AUC}
\begin{figure*}
    \centering
    \includegraphics[width=\textwidth]{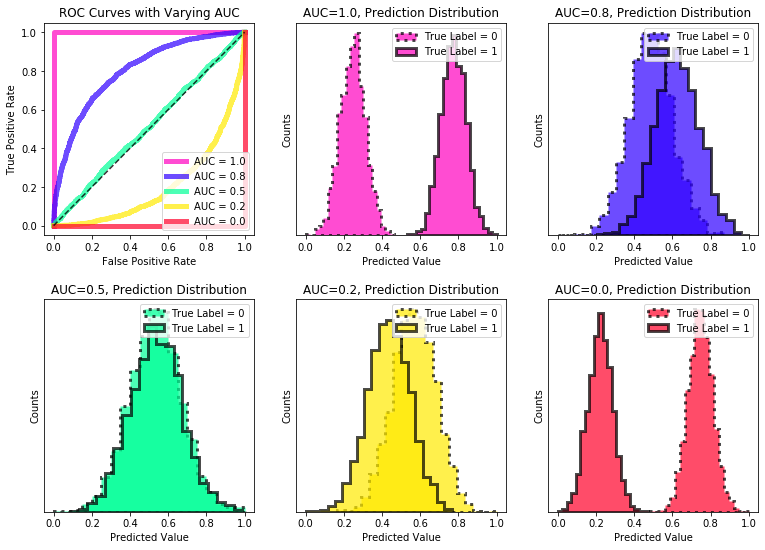}
    \caption{Visualising how AUC relates to the predictions of a given model. Each distribution relates to one ROC curve, and the AUC score is listed with each of the distributions and curves, respectively.}
    \label{fig:AUC}
\end{figure*}

Area Under Curve, or $\text{AUC}\in[0,1]$, is the area under the ROC curve. It is a measure of how well the models predictions have separated the two classes. Examples of AUC values, along with the respective distributions and ROC curves are visualised in Figure~\ref{fig:AUC}. Any value below 0.5 generally would indicate some implementation error, as the model seemingly separates the classes, but assigns the labels in reverse.

\bsp	
\label{lastpage}
\end{document}